\documentclass{article}
\usepackage{arxiv}
\usepackage{graphicx}
\usepackage[mathlines]{lineno}
\usepackage{tikz}
\usepackage{amsmath}
\usepackage{breqn}
\usepackage{tabularx} 
\usepackage{amssymb}
\usepackage{graphicx} 
\usepackage[sortcites=true,style=nature]{biblatex}
\usepackage{authblk}
\usepackage{xcolor}
\usepackage{caption}
\usepackage{siunitx}
\usepackage{upgreek, mathtools, accents}
\usepackage{setspace}
\usepackage{xr}
    
\bibliography{main}

\newcommand{\rhotot}{\rho_{\text{\tiny{tot}}}}
\newcommand{\thresh}{\rho^c_{\text{\tiny{tot}}}}
\newcommand{\cmax}{\rho_{\text{\tiny{tot}}}^{\text{\tiny{max}}}}

\definecolor{forestgreen}{rgb}{0.13,0.55,0.13}
\definecolor{orange}{rgb}{0.96,0.5,.14}
\definecolor{brown}{rgb}{0.81,0.54,.27}

\renewcommand{\eqref}[1]{Eq.~\textbf{\ref{#1}}}
\newcommand*\circled[1]{\tikz[baseline=(char.base)]{
    \node[shape=circle,draw,inner sep=1pt] (char) {#1};}}

\DeclareUnicodeCharacter{0301}{\'{e}}
\makeatletter
\newcommand*{\addFileDependency}[1]{
\typeout{(#1)}
\@addtofilelist{#1}
\IfFileExists{#1}{}{\typeout{No file #1.}}
}\makeatother

\newcommand*{\myexternaldocument}[1]{%
\externaldocument{#1}%
\addFileDependency{#1.tex}%
\addFileDependency{#1.aux}%
}
\myexternaldocument{SI}

\title{Dynamic coexistence driven by physiological transitions in microbial communities}

\title{Community states as dynamic niches in microbial communities}

\title{Community states as dynamic niches in microbial communities}

\title{Dynamic coexistence driven by physiological transitions in microbial communities}

\begin{document} 

\author[1]{Avaneesh V. Narla}
\author[1]{Terence Hwa} 
\author[2]{Arvind Murugan}

\affil[1]{Department of Physics, University of California, San Diego}
\affil[2]{Department of Physics, University of Chicago}

\onecolumn
\maketitle

\begin{abstract}
Microbial ecosystems are commonly modeled by fixed interactions between species in steady exponential growth states. However, microbes often modify their environments so strongly that they are forced out of the exponential state into stressed or non-growing states. Such dynamics are typical of ecological succession in nature and serial-dilution cycles in the laboratory. Here, we introduce a phenomenological model, the Community State model, to gain insight into the dynamic coexistence of microbes due to changes in their physiological states. Our model bypasses specific interactions (e.g., nutrient starvation, stress, aggregation) that lead to different combinations of physiological states, referred to collectively as ``community states'', and modeled by specifying the growth preference of each species along a global ecological coordinate, taken here to be the total community biomass density. We identify three key features of such dynamical communities that contrast starkly with steady-state communities: increased tolerance of community diversity to fast growth rates of species dominating different community states, enhanced community stability through staggered dominance of different species in different community states, and increased requirement on growth dominance for the inclusion of late-growing species. These features, derived explicitly for simplified models, are proposed here to be principles aiding the understanding of complex dynamical communities. Our model shifts the focus of ecosystem dynamics from bottom-up studies based on idealized inter-species interaction to top-down studies based on accessible macroscopic observables such as growth rates and total biomass density, enabling quantitative examination of community-wide characteristics. 
\end{abstract}

\section*{Introduction}

Microbial communities in natural environments are often highly dynamic~\cite{lamont2018oral,caporaso2011moving,gerber2014dynamic,david2014host,jansson2020soil,patin2018microbiome,buttigieg2018marine}. For example, many environments feature periodic replenishment of resources (e.g., the gut microbiome~\cite{gupta2022nutrient}, the ocean~\cite{schlomann2019timescales}), or resetting of other environmental factors with periods of growth between these perturbations~\cite{smith2008development,braga2018bacterial}. Lab-scale experiments ~\cite{goldford2018emergent,dalbello2021b,gowda2022genomic} on microbial ecosystems frequently adopt serial dilution cycles with dynamic environments. Recent studies have found that stable microbial communities do not settle simply into a fixed state, but are instead driven through dynamic phases involving complex changes in the environment such as depletion of oxygen and build-up of toxic waste~\cite{ratzke2018modifying,amarnath2023stress}. These changes, in turn, alter the physiological states of the microbes in the community, slowing down or even halting their growth. Changing physiological states also often change metabolic secretion and uptake profiles, and induce more complex interactions such as aggregation, motility, toxin secretion, and even contact-dependent killing~\cite{ghosh2015mechanically,rebuffat2011microcins,cao2016type,garcia2017contact}. 

These observations have been interpreted using models from theoretical ecology that typically explain ecosystem assembly and stability~\cite{may1973stability, May+2001} in terms of resource competition~\cite{newton2023modulation,goldford2018emergent,dalbello2021b}, niche differentiation~\cite{nuccio2020niche} and competitive exclusion~\cite{hardin1960competitive}. However, these models typically assume that communities and the organisms in them are at steady state ~\cite{macarthur1967limiting,macarthur1970species,tilman1977resource,may1973stability,volterra1928variations}. 
This difference between empirical observations and theoretical models raises questions about the role of dynamic physiological state changes in forming complex communities. One possibility is that physiological state changes are merely details, not essential for understanding factors that enable community assembly. In this perspective, nothing is lost by coarse-graining over dynamics and modeling communities as if they are at steady state. Microbial communities would be expected to show similar complexity and structures if microbes stay in fixed states (e.g., exponential growth) and independent of whether interactions through metabolic secretion and uptake occur in a temporally staged manner.  

Another possibility is that physiological state changes create dynamic niches that support complex communities. Since microbes have a plethora of non-growing states, this scenario could significantly expand the ways of generating niches beyond well-studied cases such as distinct metabolites~\cite{tilman1977resource}, space~\cite{harrison2010spatial}, and externally dictated temporal epochs (e.g., diel or annual cycles)~\cite{kronfeld2003partitioning}. Further, the nature of such self-generated dynamic niches, if they exist, might have signatures that are predicted to be observed in microbial communities. 


We cannot easily address this question about the role of state changes using the current bottom-up theoretical frameworks (e.g., Lotka-Volterra or Consumer-Resource models) since these models typically characterize organisms and their interactions with fixed parameters. In these models, community dynamics only involves changes in species abundances and nutrient concentrations and is justified by assuming organisms are in a fixed physiological state, (e.g., Monod growth for exponentially-growing cells~\cite{monod1949growth}).  If one is to adopt a model of interactions between each species and its environment (as in Consumer-Resource Models) or other species (as in Lotka-Volterra Models), then each physiological state would minimally involve a different set of uptake and excretion parameters; a given species would effectively be modeled as multiple species over time. Thus, bottom-up models of dynamic communities require extensive characterization and unconstrained assumptions on specific details about what different cells do in different conditions.

As a first step towards quantitatively modeling communities of species that undergo physiological changes, we introduce a minimal top-down phenomenological framework, the Community State Model. Our model is phenomenological at the level of species density; the physiological state and thus the growth rate of each species in a community is assumed to depend only on the community biomass at any given time, and as a result, community states are defined by regions of biomass density. Such a model can be solved explicitly (numerically and in simple cases analytically) to yield the temporal organization of community dynamics at a quantitative level. 

Analysis of the Community State model points to sequential dynamics as a strategy to form a stable community involving a large number of species~\cite{horn1974ecology}. In this simple model, each species grows rapidly in one (or a few) community states that persist over specific intervals of biomass accumulation, with slower or even no growth in other parts of the inter-dilution period (hereby referred to as the growth period). This strategy is a distinct alternative to the co-growth strategy based on steady-state models with fixed physiological states where species grow on resource niches simultaneously. 

For this sequential coexistence strategy, our model allows us to uncover a number of key features of community dynamics. We find (a) tolerance of community diversity to fast-growing species if such growth is limited to specific community states, (b) enhanced community stability through staggered dominance of different species in different community states, and (c) a requirement of increased growth dominance for late-growing species. These features counteract the dominant notions regarding species competition derived from analysis of steady-state systems, and serve as principles to guide the understanding of complex dynamical ecosystems


\begin{figure*}
    \centering
    \includegraphics[width=\columnwidth]{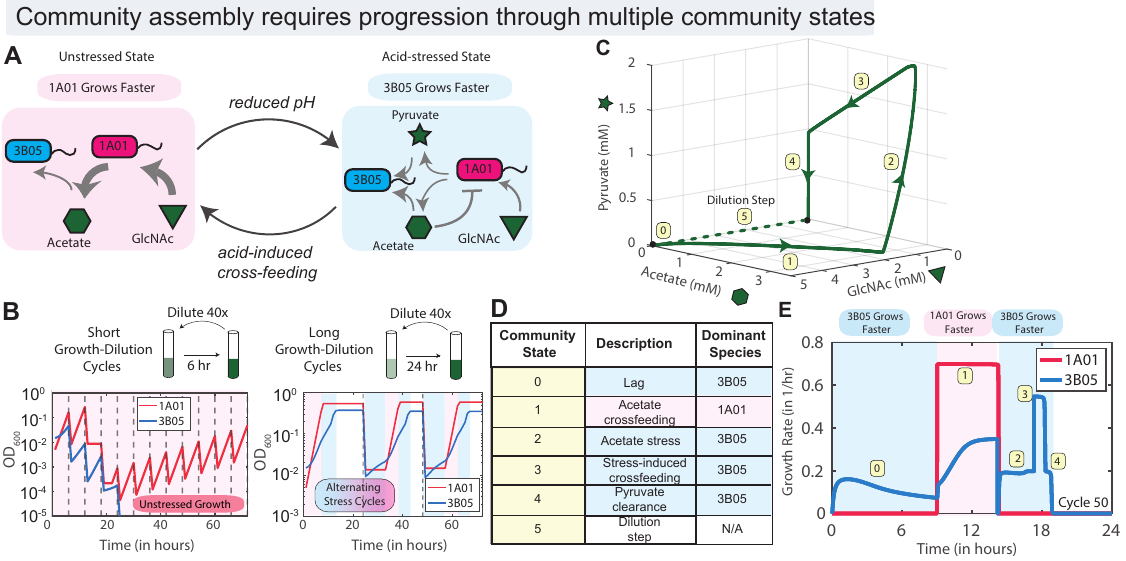}
    \caption{
    \textbf{Sequential transitions in a community's state revealed in a simple co-culture of a sugar consumer and an organic acid consumer.}
    (A) Schematics describing the dynamics of a co-culture of marine bacteria {\em Vibrio\ sp.~1A01}, a sugar consumer, and {\em Neptumonas\ sp.~3B05}, an organic acid consumer, growing on N-acetyl glucosamine (GlcNAc) under repeated serial-dilution cycles: the co-culture passes from an initial unstressed state in which 1A01 grows faster than 3B05 (pink box), to an acid-stressed state in which 3B05 grows but 1A01 does not (blue box). 
    Pointed gray arrows indicate metabolic flow (thickness indicates flow magnitude) and blunt-end arrows indicate growth inhibition. (B) Experimental investigation by Amarnath et al.~\cite{amarnath2023stress} revealed coexistence only if serial-dilution cycles were sufficiently long to allow for an intricate sequence of `community states', i.e.,  different combinations of the physiological states of each species and media conditions, labeled by the numbers in (C,D). (C) shows changes in 3 major metabolites while Table (D) describes the community states along the green path in (C). (E) The growth rate of each species along the path in (C), i.e., between two serial dilutions (steady state cycle shown).}
    \label{fig:kapilbasics}
\end{figure*}

\begin{figure*}
    \centering
    \includegraphics[width=\columnwidth]{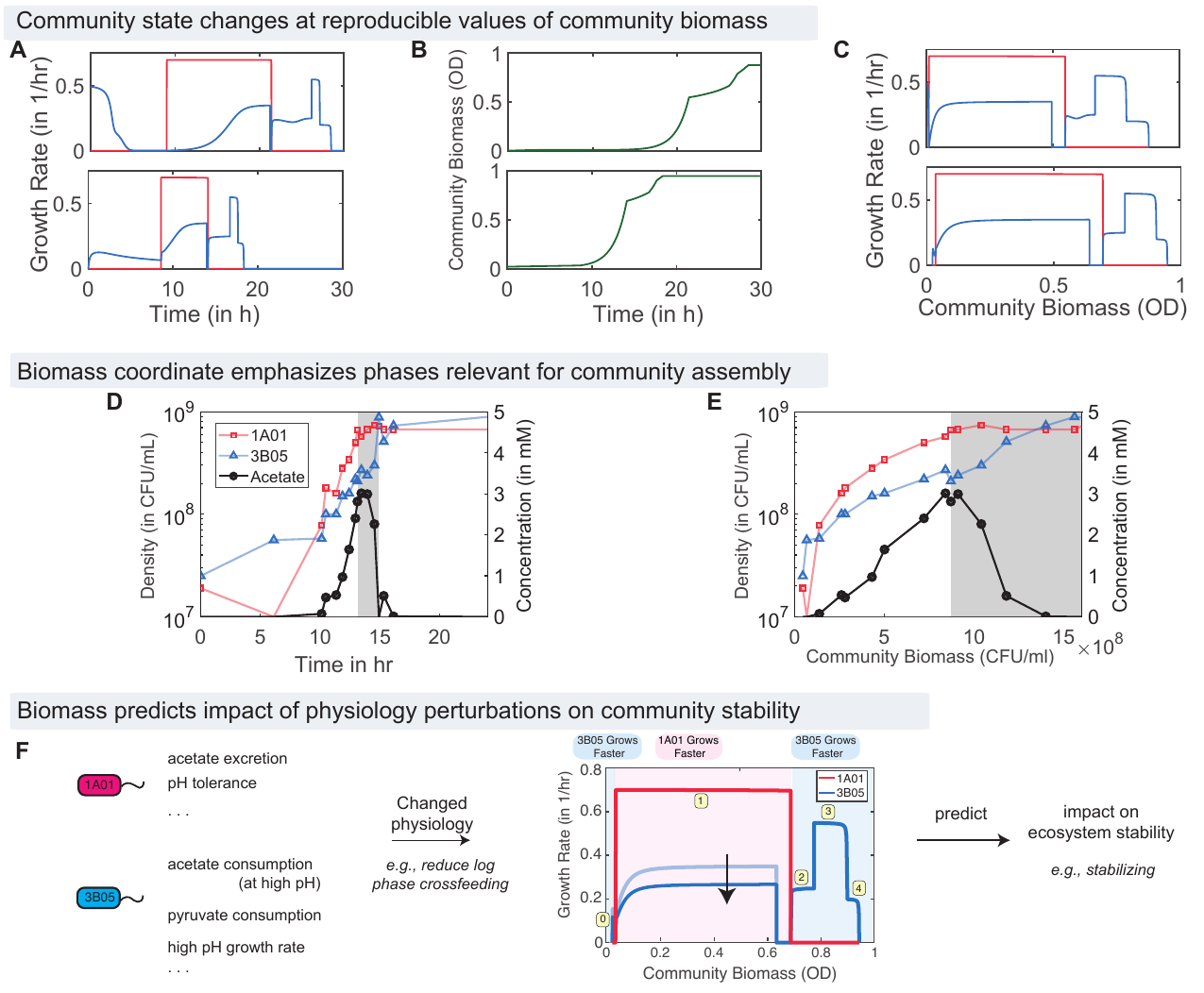}
    \caption{\textbf{Benefits of indexing community states by the accumulated community biomass.} 
     When plotted as a function of time during a growth period, different initial conditions (here, relative species abundances of A:B=1 for top row, A:B=0.66 for bottom row) for the system in Fig.~\ref{fig:kapilbasics} lead to (A) different growth rate curves of each species and (B) accumulated community biomass density. (C) However, growth rate data as a function of community biomass density is relatively reproducible. Results shown from second cycle (before reaching stable cycle). 
     (D,E) Experimental data from~\cite{amarnath2023stress} on cell densities and acetate concentration during a growth dilution cycle. (D) The most important physiological changes that underpin coexistence in~\cite{amarnath2023stress} occur in a short window of time (gray) where acid stress causes dramatic changes in acetate concentration (black curve) and other metabolites (not shown). However, without prior knowledge of the acetate stress mechanism, sampling uniformly in real-time will dedicate many time points to the lag phase (relatively unimportant for coexistence~\cite{amarnath2023stress}) and may miss the critical gray region. (E) In contrast, sampling evenly in community biomass naturally emphasizes the gray region. Thus investigating metabolites that change dramatically between biomass density intervals can assist in identifying the mechanistic basis of community assembly. (F) A mutation that changes physiology in specific environments (here, reducing log phase crossfeeding) will change growth curves in community biomass intervals corresponding to those environments (here, the light blue growth curve for 3B05 is lowered to the dark blue curve in the pink biomass interval). In this work, we derive a formula relating growth rate curves as a function of biomass to coexistence. Consequently, our top-down framework can relate changes in physiological properties to community assembly.}
    \label{fig:kapilbiomass}
\end{figure*}

\section*{Results}
\subsection*{Case study of community dynamics during serial-dilution cycles}
\noindent 
The model of microbial community dynamics developed here is inspired by dynamics revealed by a recent study of a seemingly simple cross-feeding system subjected to repeated serial-dilution cycles~\cite{amarnath2023stress}: Two species of marine bacteria, \textit{Vibrio sp.~1A01} and \textit{Neptunomonas sp.~3B05}, isolated from a chitin-degrading coastal community~\cite{datta2016microbial}, were grown on N-acetyl glucosamine (GlcNAc) as the sole carbon and nitrogen sources in batch culture. 1A01 consumes GlcNAc and excretes acetate and ammonium, while 3B05, which does not consume GlcNAc, can grow on the acetate and ammonium excreted by 1A01 (Fig.~\ref{fig:kapilbasics}A). Because 1A01 grows on GlcNAc faster than 3B05 grows on acetate, acetate inevitably accumulates in the medium, becoming toxic when reaching the buffer capacity of the medium (which is low for seawater). In canonical syntrophy, toxicity would slow down the toxin-excreting species more than the toxin-clearing species, resulting in a stable state where both species grow exponentially~\cite{estrela2012metabolism}. This is not the case for 1A01-3B05 co-culture: acetate accumulation in the environment slows down the growth of 3B05, the acetate-consumer, more than that of 1A01, the acetate excretor. Thus acetate is expected to accumulate, with the arrest of the co-culture once acetate exceeds the buffer capacity. Yet, as shown in Fig.~\ref{fig:kapilbasics}B, when this system was subjected to 24-h growth-dilution cycles, the system miraculously cured itself of acetate accumulation, with the two species reaching a stable, comparable abundance ratio according to samples taken at the end of each cycle after a few cycles. 

Detailed analysis of the two-species dynamics revealed that stability was achieved through physiological transitions during the 24-h growth period, as shown in Fig.~\ref{fig:kapilbasics}C-E: 
The first phase \circled{0} after nutrient replenishment is a lag phase for 1A01; in the next phase \circled{1}, acetate accumulated and pH dropped; 1A01 grew fast with slower growth for 3B05. 
Phase \circled{2} commenced when the pH hit a critical threshold (set by the pKa of acetic acid) which caused 1A01 and 3B05 to enter growth arrest, with 1A01 excreting large amounts of glycolytic intermediates (e.g., pyruvate). This stress-induced excretion played a key role in stimulating the growth of 3B05 (phase \circled{3}), with the consequence of  
removing acetate from the medium and restoring pH (phase \circled{4}). This in turn allows 1A01 to avoid death and be available for dilution into the next cycle (phase \circled{5}). 

Amarnath et al.\cite{amarnath2023stress} showed that this highly dynamical mode of coexistence is not specific to the marine species studied: dynamic coexistence through similar acid shock and recovery was shown also for co-culture of species taken from a soil community or even between enteric and soil bacterium. 
Metabolic analysis in~\cite{amarnath2023stress,taylor2022metabolic} suggests that such interactions are generic between species with complementary sugar-preferring vs.\ acid-preferring bacteria, or between glycolytically-oriented vs.\ gluconeogenically-oriented modes of metabolism. Thus, dynamic coexistence with each species passing through multiple physiological states in a cycle may be the norm rather than the exception~\cite{mccully2018escherichia,estrela2022functional,goldford2018emergent,li2018hidden}. The lack of reports of such dynamical features may reflect the lack of time-resolved measurements, which occurred within a few-hour window of the 24-hour growth-dilution cycles. The focus of existing theoretical studies in ecology on steady-state characteristics and stable coexistence of many species~\cite{macarthur1964competition,May+2001} may also contribute to the lack of measurements on dynamic characteristics. 

\subsection*{Benefits of indexing community dynamics by community biomass}

How can we effectively capture the key drivers of the complex dynamics underlying this system? We propose a phenomenological and experimentally accessible description, using ``community biomass'' as a proxy for key drivers of community dynamics. 

One advantage of this biomass-based description is that the community state changes at reproducible values of the accumulated biomass for perturbations in external parameters. This robustness arises because changes in environmental parameters like pH and oxygen levels are typically accompanied by biomass accumulation. In contrast, the most direct description based on time has several drawbacks. As shown in Fig.~\ref{fig:kapilbiomass}A, community dynamics during a growth cycle in real time are highly variable as initial species or nutrient abundances are varied. However, this variance is mostly counteracted by the variation in the real-time accumulation of biomass; Fig.\ref{fig:kapilbiomass}B. Hence, combining Fig.~\ref{fig:kapilbiomass}A and \ref{fig:kapilbiomass}B, we find that the growth rate as a function of accumulated biomass is relatively reproducible; see Fig.\ref{fig:kapilbiomass}C. See \textit{Supplementary Supplementary Figures S1-2} for other ecosystems with even higher reproducibility of biomass. While biomass values corresponding to community state changes will be different when, say, a given species is part of a novel community, these results suggest that total biomass could be part of a useful top-down description of a given ecosystem. 

Another advantage of indexing the community dynamics by total biomass is that it naturally emphasizes growth phases most relevant for community assembly. As shown in Fig.~\ref{fig:kapilbiomass}D, when plotted in real time, the most critical physiological changes occur in a relatively brief period indicated by the gray band (where acetate buildup hits a threshold, leading to subsequent stress-induced crossfeeding of pyruvates and other metabolites). There would have been no reason to sample the short time period represented by the gray band in Fig.~\ref{fig:kapilbiomass}D, without data on acetate (black curve) and other results of the detailed mechanistic study in~\cite{amarnath2023stress}. More generally, the drawback of real time is because environmental change is a result of \emph{absolute} biomass growth, and for an exponentially-growing culture, the same absolute amount of environmental change takes exponentially less time as the culture grows. Instead, sampling the ecosystem uniformly in accumulated community biomass density (Fig.~\ref{fig:kapilbiomass}E) emphasizes important periods such as the gray region, even if the role of acid stress was unknown. 

Finally, as we will detail later, the accumulation of sufficient biomass is the necessary and sufficient requirement for each species to be maintained in stable cycles. Consequently, we can use this picture based on biomass to predict the impact of, e.g., altered physiology in an organism due to mutations, on coexistence. For example, consider a change that decreases cross-feeding during exponential growth as shown in Fig.~\ref{fig:kapilbiomass}F. Naively, such a change would be expected to destabilize coexistence as cross-feeding is believed to enhance coexistence~\cite{goldford2018emergent}. But our biomass-based results derived below will predict the opposite; the depicted physiological change causes 3B05 will grow slower during 1A01's growth phase. 
We will show that reducing co-growth in intervals of the biomass coordinate will generally favor coexistence since each species will get guaranteed (but capped) growth in the community state that it dominates. Consequently, our framework will predict reducing cross-feeding during the exponential phase can stabilize the community, counter to intuition.

\subsection*{A top-down model for complex communities}
\begin{figure}
    \centering
    \includegraphics[width=\columnwidth]{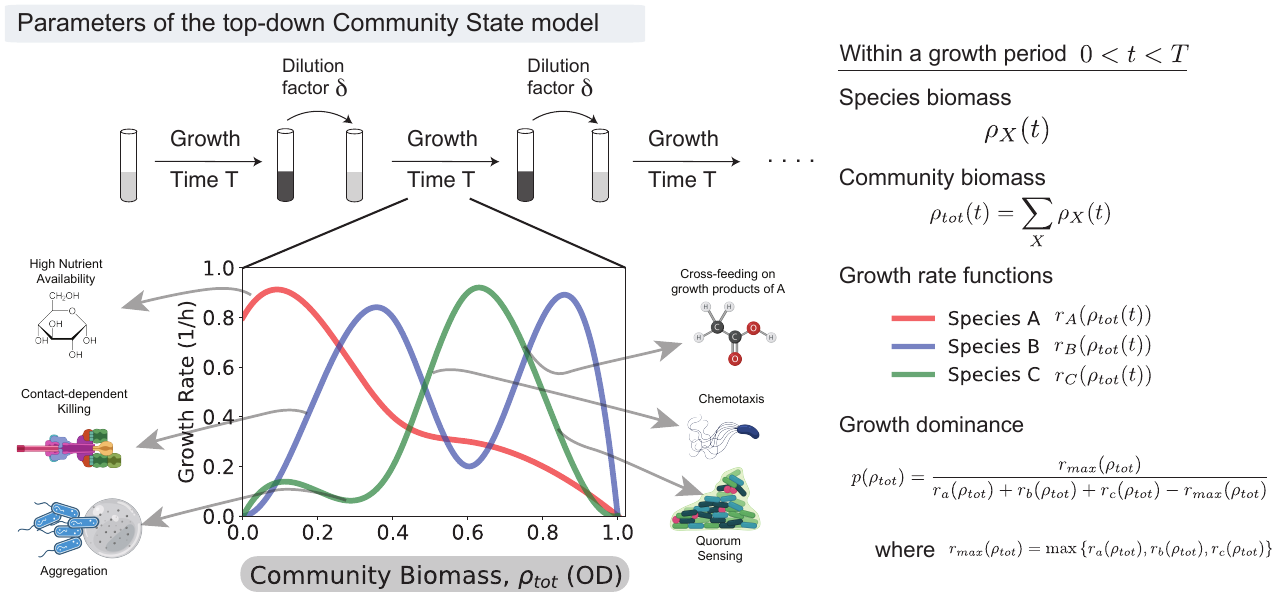}
    \caption{\textbf{Top-down description of ecosystems in the Community State Model.} We consider ecosystems that grow, gaining biomass for a period $T$, before being diluted into a fresh environment. During the growth period,
    the community changes its environment through many complex processes shown, leading to a sequence of community states. The Community State model assumes these processes to be turned on and off at different points along a one-dimensional phenomenological coordinate that parameterizes the sequence of community states. In the simplest version of the model shown, this coordinate is taken to be the accumulated community biomass $\rho_{tot}(t)$ since nutrient depletion, buildup of toxins, and spatial structure, etc. are accompanied by biomass growth. Each species is assigned a different set of growth rates $r_X(\rho_{tot})$. At any given stage of the growth cycle, indexed by $\rho_{tot}$, growth dominance $p(\rho_{tot})$ is the ratio of the growth rate of the fastest-growing species to that of other species. As suggested by the cartoon, different species might dominate at different $\rho_{tot}$ and to different extents $p(\rho_{tot})$.}
    \label{fig:phenomodel}
\end{figure}

\noindent We propose a general model - the Community State (CS) Model - for investigating multi-species dynamics in microbial communities in cyclic environments. In this top-down model, we take the growth rate, $r_\alpha(S)$, of each species $\alpha$ to depend on the community state $S$, which progresses through multiple states due to various environmental changes driven by the microbes themselves as shown in Fig.~\ref{fig:phenomodel}. In the simplest model, we assume that the community state $S$ can be parameterized by the community biomass, taken to be the total cell density $\rhotot^{(j)}(t)\equiv\sum_\alpha \rho^{(j)}_\alpha(t)$, where $\rho^{(j)}_\alpha(t)$ is the cell density of species $\alpha$ at time $t$ in the $j^{\rm th}$ cycle. (Here, we assume all species to have the same biomass per cell. More generally, a scaling factor can be introduced to absorb the species-dependent cell mass.) Thus, the growth of species during the $j^{\rm th}$ cycle is described by
\begin{equation}
    \frac{d}{dt}\rho^{(j)}_\alpha=r_\alpha\left(\rhotot^{(j)}(t)\right)\cdot \rho^{(j)}_\alpha(t) \qquad {\rm for }\ 0\le t \le T. \label{dynamics}
\end{equation}
When $t$ reaches the growth period $T$, all densities are reduced by a common factor $\delta<1$,  i.e., 
\begin{equation}
\rho^{(j+1)}_\alpha(t=0)=\delta \cdot \rho^{(j)}_\alpha(T). \label{dilution}
\end{equation}
Eqs.~[\ref{dynamics}] and [\ref{dilution}] define an effective ``map'' for the density of each species at the beginning of each cycle, $\rho^{(j)}_\alpha(0)$ starting from the initial composition $\rho^{(1)}_\alpha(0)$. For convenience, we choose the growth period $T$ to be sufficiently long such that the total cell density has time to reach the maximum $\rhotot^{\rm max}$ where $r_\alpha = 0$ for all species. Thus, the dynamics of this model are governed by the growth rate functions $r_\alpha(\rhotot)$, the maximal cell density of the system $\rhotot^{\rm max}$, and the dilution factor $\delta<1$, independent of the growth period $T$. 

\begin{figure*}[!ht]
    \includegraphics[width=\columnwidth]{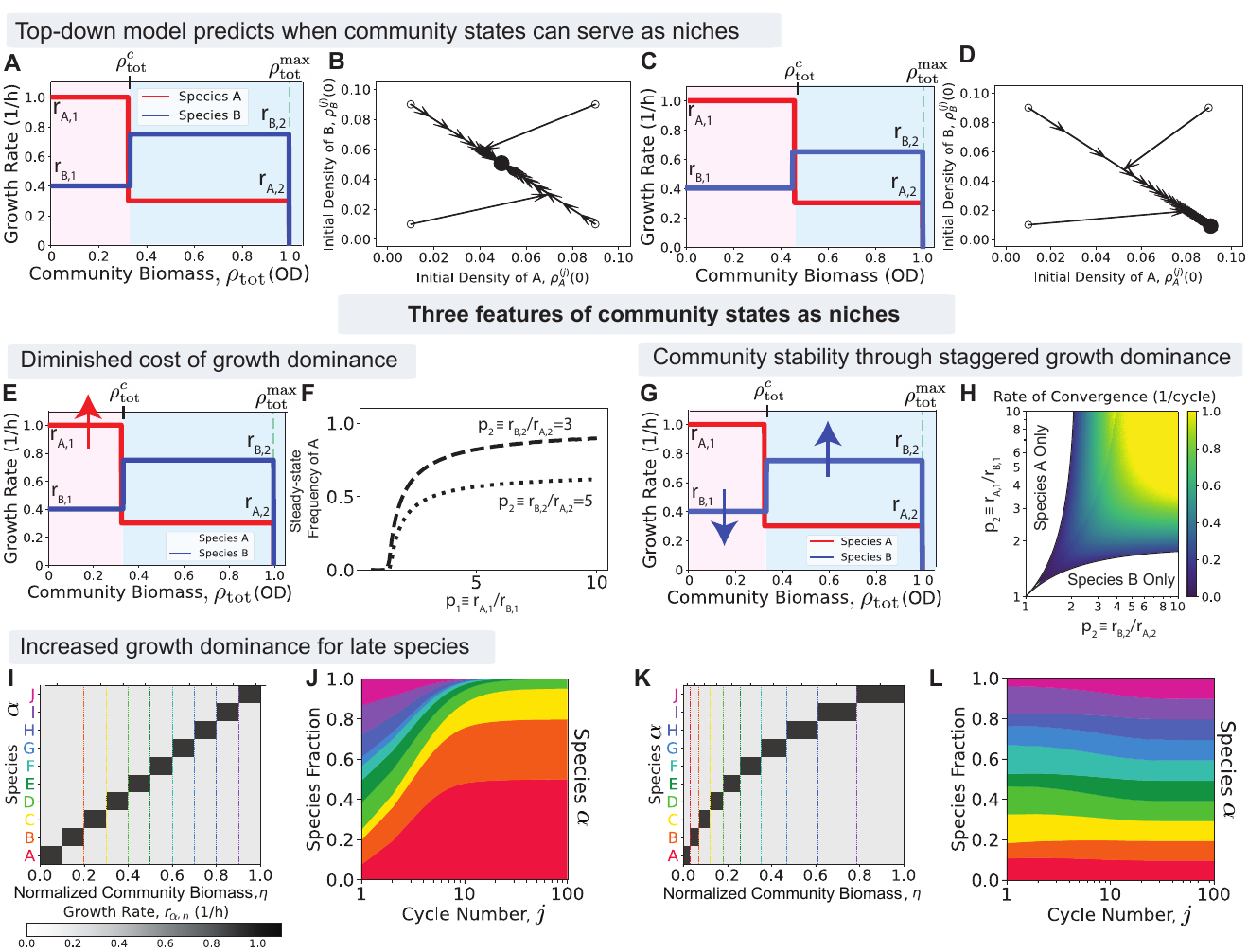}
    \caption{
    \textbf{Key features of community states as niches}
    (A) Growth curves $r_X(\rho_{tot})$ for two species; species $A$ grows rapidly in the early community state that lasts for community biomass $\rho_{tot} < \rho_c$; species $B$ grows faster when $\rho_{tot} > \rho_c$.  (B) Species densities at the end of consecutive serial dilutions from different initial conditions (black open circles) for growth curves in (A) converge to a stable coexistence point (black solid circle). (C,D) Same as (A,B) but the modified growth curves in (C) lead to extinction of species $A$ shown in (D). While growth curves in (A),(C) are visually similar, Eq.\ref{eqn:toyAB_invasion},\ref{eqn:toyBA_invasion} predict that the two community states can support two species in (A) but not in (C). \textbf{Key features:}
    (E) We increase growth dominance $p_1 = r_{A,1}/r_{B,1}$ of species $A$ in the first community state. (F) Steady-state abundance of species $A$ as a function of growth dominance $p_1$ shows a quickly saturating effect.
    (G) We consider a coordinated change of growth dominance $p_1,p_2$ across multiple community states. (H) Rate of convergence back to stable coexistence point (i.e., Lyapunov exponent of stability) as a function of growth dominances $p_1,p_2$; ecosystem stability increases without saturation for coordinated changes in dominance. 
    (I) A 10-species ecosystem with each species ``dominant'' with high growth rate $r_+$ in a unique community state and $r_-$ elsewhere; all community states last equal intervals of normalized community biomass $\eta \equiv (\rhotot/\cmax-\delta)/(1-\delta)$. (Here $\delta$ is the dilution factor between growth cycles. Growth dominance $p = r_+/r_- = 5$.)
    (J) Stacked chart shows species abundance at the end of each cycle (in fraction of total biomass) over multiple serial-dilution cycles. (K,L) Same as (I,J) but with wider biomass intervals for later community states (see Eq.\ref{eq:delta-eta}). The wider intervals for later species compensate for the priority effect enjoyed by early species.}
    \label{fig:threeresults}
\end{figure*}

The assumption that growth rates depend only on accumulated community biomass can be justified mechanistically in some cases; see \textit{Supplementary Text S2.2 and Supplementary Figures S1-2}. But from a phenomenological point of view, this assumption can be viewed as the minimal closure of the equations describing an ecosystem that results in a coexistence criterion. Such a coexistence criterion is derived in \textit{Supplementary Text S2.6} for mutual invasibility between two species. E.g., for the ability of one species, $A$, to invade a monoculture of species $B$, we find
\begin{equation}
    I_{A,B} \equiv \int_{\delta\cdot\cmax}^\cmax \frac{r_A(\rhotot)}{r_B(\rhotot)}\cdot 
    \frac{d\rhotot}{\rhotot} > \log \left(1/\delta\right).\label{eqn:IAB}
\end{equation}
Similarly, the invasion of a species $A$ monoculture by species $B$ requires $I_{B,A} > \log \left(1/\delta\right)$.

Intuitively, these conditions suggest that coexistence imposes conditions on the relative growth rate $r_A(\rho_{tot})/r_B(\rho_{tot})$ in different intervals of community biomass $\rho_{tot}$. For example, consider the growth curves $r_A(\rho_{tot}),r_B(\rho_{tot})$ shown in Fig.~\ref{fig:threeresults}A. Species $A$ grows faster than $B$ for low biomass $\rho_{tot} < \rho_c$, with a growth dominance $p_1 = r_{A,1}/r_{B,1}$. Conversely, species $B$ grows faster for higher biomass $\rho_{tot} > \rho_c$, with growth dominance $p_2 = r_{B,2}/r_{A,2}$. Applying \eqref{eqn:IAB} to this toy model with step-like growth curves as shown in Fig.~\ref{fig:threeresults}A and C, we find conditions for mutual invasibility and thus for stable coexistence:
\begin{align}
 p_1\log (\thresh/(\delta\cdot\cmax))& > \log 1/\delta -\frac{\log (\cmax/\thresh)}{p_2}, \label{eqn:toyAB_invasion}\\
 p_2\log (\cmax/\thresh)& > \log 1/\delta -\frac{\log (\thresh/(\delta\cdot\cmax))}{p_1}. \label{eqn:toyBA_invasion}
\end{align}
When the growth dominances $p_1,p_2$ and the width of biomass interval $\thresh$ satisfy these conditions (\eqref{eqn:toyAB_invasion}, \eqref{eqn:toyBA_invasion}), serial dilution cycles starting from any initial condition converge to a unique fixed point where both species are present; see Fig.~\ref{fig:threeresults}B. Outside of this regime, e.g., with visually similar growth curves shown in Fig.~\ref{fig:threeresults}C, one of the two species goes extinct over multiple cycles as shown in Fig.~\ref{fig:threeresults}D. See \textit{Supplementary Text S2.4} for more analyses.

It is tempting to interpret community states in the two intervals of biomass shown in Fig.~\ref{fig:threeresults}A as two dynamic `niches' where species $A$ and $B$ dominate respectively, thus guaranteeing their coexistence. However, as the distinct steady-state compositions for similar growth curves in Fig.~\ref{fig:threeresults}A and C show (coexistence and competitive exclusion respectively), community states can serve as distinct niches that support distinct species only under specific conditions. Thus the nature of niches that arise from community states remains unclear -- what is the impact of the width of biomass intervals over which a community state persists, what effect does their temporal ordering have, and can community states support complex communities with many species? The phenomenological Community State model can be used to address these questions and makes non-trivial predictions on how growth and non-growth states must be structured across a community for stable coexistence. Below, we highlight three insights derived from these equations into the nature of niches based on community states:

\noindent \textbf{Tolerance to growth dominance:}
As shown in Fig.~\ref{fig:threeresults}E and F, the growth dominance $p_1$ of species $A$ in the first biomass interval negatively impacts the steady state coexistence ratio at small $p_1$ but saturates at larger values, being limited by the value of the other growth dominance, $p_2$. This diminishing damage to coexistence by a species with strong growth dominance can also be quantified by the tolerance range of parameters like $\rho_c$ that allows for coexistence (see Supplementary Figure S3), follows from the structure of \eqref{eqn:IAB}.



The tolerance to growth dominance contrasts sharply with one of the most basic tenets of ecology, that faster-growing species drive slower ones to extinction, which is at the root of the ``paradox of the plankton'' \cite{tilman1977resource}. It allows individual species with significant growth advantages to coexist with other slower species (thus retaining ``services'' by the latter in other more challenging community states), provided that these advantages are limited to some specific physiological states. This effect plays a pivotal role in the coexistence of larger communities to be described below.

\noindent \textbf{Community stability through staggered growth dominance:}
While coexistence tolerates strong individual growth dominances as described above, the coexistence regime is broadened if \emph{both} $p_1$ and $p_2$ are large, i.e., if the species stagger their dominance in distinct community states. The impact of staggered dominance is seen also in the robustness of coexistence, measured by the convergence rate to the stable cycle following small perturbations (the Lyapunov exponent); see Fig.~\ref{fig:threeresults}G, H. This metric is also a measure of the stability of the ecosystem against environmental or physiological fluctuations. 

Since enhancement in the size and stability of the coexistence region requires increases in growth dominance in {\em distinct} community states (i.e., $p_1$ and $p_2$), such a communal effect is aided by different species coordinating their growth dominance across multiple community state. However, individual species can also contribute to such staggered growth dominance since a species can increase the growth dominance of another species in another community state by reducing its own growth rate in that state; e.g., species $A$ can increase $p_2$, the dominance of species $B$ in state 2, by decreasing $r_{A,2}$. Thus, this global coordination is facilitated by individual species {\em specialized} to dominate in distinct community states.

\noindent\textbf{Increased growth dominance for late species:} 
The coexistence criteria $I_{A,B},I_{B,A}$ are not symmetric between species $A$ and $B$. This asymmetry can be traced to a ``priority effect'' where species $A$ capitalizes on early growth, accumulating large numbers while species $B$'s growth occurs in a later biomass interval~\cite{wang2021complementary,erez2020nutrient,fukami2015historical}. The consequences of this priority effect are shown for a $N$-species community in Fig.~\ref{fig:threeresults}I,J: if each species is dominant in a unique biomass interval of equal width (with growth dominance $p=5$ in each community state), late-growing species are driven extinct after several cycles. 

We find that members of such a community can all coexist despite the priority effect if early-growing species occupy narrower biomass intervals. Based on \eqref{eqn:toyAB_invasion} and \eqref{eqn:toyBA_invasion}, we were able to derive a special distribution of biomass interval (see \textit{Supplementary Text S2.8}) for coexistence. Expressed in the normalized biomass coordinate $\eta = \rho_{tot}/\rho_{max}$, if the width interval $\Delta \eta_n$ for the species growing in the $n^{\rm th}$-state follows an exponential distribution
\begin{equation}
    \Delta \eta_n\sim (1/\delta)^{n/(N+p-1)},  \label{eq:delta-eta}
\end{equation}
then late species coexist with early species and are, in fact, equi-abundant (provided that growth dominance $p$ does not greatly exceed $N$); see Fig.~\ref{fig:threeresults}K, L. Thus, an appropriate choice of transition points between community states favoring the late-growing species can counteract the priority effect. Alternatively, the priority effect can be counteracted by stronger growth dominance of the late-growing species even if the biomass intervals of the different states are equal; see Supplementary Figure S6. These results underscore and quantify the large (exponentially increasing) burden faced by the late-growing species to be included stably in large communities.

\begin{figure*}
    \centering
    \includegraphics[width=\columnwidth]{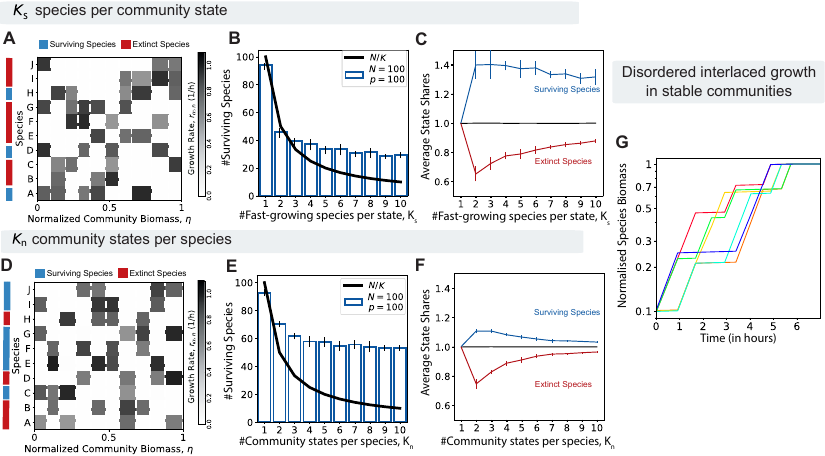}
    \caption{
    \textbf{Competition and coexistence in complex communities.} (A) $K_s$ model of random preferences: $K_s$ randomly chosen species are given high growth rate $r_+$ in each community state that lasts for a fixed interval of community biomass $\eta$; all other species are set to slow growth $r_-$ in that interval.  Illustrative example shown in (A) with $K_s = 4$, $N=10$ total species and growth dominance $p = r_+/r_- = 100$. 
    After many serial-dilution cycles, some species persist (blue) while others go extinct (red).
    (B) Average number of surviving species with $N=100$ exceeds naive expectation $N/K_s$ (black line). (C) The average number of `state shares' for surviving species and extinct species as a function of $K_s$; here, `state share' is the share of community states that each species grows quickly in. For example, Species A in Fig. 5A receives 1/4 of each state in grows quickly as it shares the fast growth with three other species, and thus has a total of 1.25 state shares.
    (D) $K_n$ model of random preferences: Each species is assigned a high growth rate $r_+$ in $K_n$ randomly chosen community states and slow growth $r_-$ in other states. Illustrative example shown with $K_n=4, N= 10$. (E,F) Same as (B,C) but for the $K_n$ model. (F) Average number of `state shares' for surviving species and extinct species as a function of $K_s$; `state share' is defined as in (C), but here each species may share each state with a different number of species.
     (G) Time course of normalized biomass for the surviving species over a stable cycle in an illustrative example of 10 species with $K_s = 4$ preferred niches and $K_n = 4$ competitors in each niche (imposing constraints of both  $K_s$ and $K_n$ models). Species show a complex disordered pattern of growth and no-growth states with subtle correlations that enable coexistence.
     (Error bars in all panels indicate standard deviation from 10 simulations of different growth matrices with $30\%$ variation in $r_+$, $r_-$, and $\Delta\eta$.)}
    \label{fig:off_diag}
\end{figure*}

\subsection*{Complex communities}

We next consider a multispecies Community State model in which different species can grow (and hence compete) in the same community state. This model can be represented by a ``preference matrix'' where each element of the matrix denotes the community states in which each species grows preferentially and the boundaries of the community states are common to all species and given by defined biomass values ($\eta_{i-1}$ and $\eta_{i}$ for the ith community state). We allow for multiple species to grow preferentially in any community state. This structure allows us to see the effect of competition between species growing preferentially in the same community states.

We first explored the case where each community state can support the rapid growth of a fixed number $K_s$ of species. An example of a preference matrix is shown in Fig.~\ref{fig:off_diag}A; each community state was assumed to last for the same average interval $\Delta \eta$ of biomass. In each community state (i.e., column of matrix shown), we assigned a high growth rate $r_+$ for $K_s$ randomly chosen species and a low growth rate $r_-$ that is a $p=100$ times lower than $r_+$ for all other species. $30\%$ random variation added to each parameter; see Supplementary Information for other details. The results, shown in  Fig.~\ref{fig:off_diag}B), show an expected decrease in steady-state diversity with an increasing number of species $K_s$ competing in each community state. However, the decrease in surviving species is slower than a naive expectation of $N/K_s$.

To gain insight into which species survive, contrast rows of the matrix corresponding to surviving (blue) and extinct species (red) in Fig.~\ref{fig:off_diag}A. Survivors (blue) tend to grow fast in five or more community states while extinct species (red) have mostly three preferred states. To test this hypothesis, we plotted the average number of preferred states (states in which a species grows at $r_+$) for the surviving and extinct species for the $N=100$ system (Fig.~\ref{fig:off_diag}C); we see that the surviving species are generalists who can grow rapidly in multiple community states. 

We next explore the situation where each species has a fixed number $K_n$ of preferred states, i.e., now, all species are generalists to an equal extent. However, the identity of the preferred community states for each species is randomly chosen; see Fig.~\ref{fig:off_diag}D for an example matrix. The number of surviving species is now clearly increased compared to the $K_s$ model; compare Fig.~\ref{fig:off_diag}E to Fig.~\ref{fig:off_diag}B. What factors distinguish the surviving and extinct species? The specific example in Fig.~\ref{fig:off_diag}D suggests that extinct species had more competitors in their preferred community states. To confirm this, we calculated the average number of preferred states for each species, weighted by the number of competitors that also grow fast in that state. Plotting this competition-weighted average state share for the $N=100$ system, we find the surviving species (blue) indeed have a higher share of their preferred community states compared to extinct species (red). See Fig.~\ref{fig:off_diag}F. 

In Supplementary Figure S7 and S8 and Supplementary Text S2.9, we consider a Consumer Resource model in a chemostat with each species growing on multiple resources. We explored similar constraints as above, with the total number of species growing on each resource ($K_s$) fixed or the number of resources each species grew on ($K_n$) fixed. We find that the fraction of surviving species for fixed $K_n$ (43\% for $K_n=10$) is lower than in the CS Model (52\% for $K_n=10$) and higher for fixed $K_s$ (34\% for $K_s=10$) than in the CS Model (28\% for $K_s=10$).

\section*{Discussion}

In this work, we have investigated models of growth and survival of microbial species in communities subjected to cyclic environmental fluctuations, focusing on the case of prolonged periods between nutrient replenishment as seen often in the wild~\cite{datta2016microbial,kolter1993stationary}; under these conditions, exponential steady-state growth cannot be sustained. In the lab, non-steady-state growth can occur during serial-dilution cycles where the cycle length is long enough for nutrient depletion or build up for toxic waste that limits growth~\cite{goldford2018emergent,smith2011bacterial,taylor2022metabolic}. 
Accurate bottom-up models are not feasible given our limited understanding of microbial behaviors outside of exponential growth ~\cite{kolter1993stationary,navarro2010stationary}. 

Inspired by a model experimental system, we investigated the creation of dynamical niches by a combination of physiological states taken by members of a community in response to self-generated environmental changes. 
In our model, each species $X$ is assigned a growth rate $r_X(S)$ in each state $S$ of the community. 
A community state $S$ was taken to last for an interval of total community biomass $\rho_{tot}$ accumulated during the growth of the community, and the corresponding growth rate of each species in that community state $S$ reflects the physiological state each species is in as well as its environmental context (which includes the physiological states of all other species). Using total community biomass $\rho_{tot}$ as a driver of community state transitions gives 
a simple model that allows us to derive quantitative self-consistency conditions on temporal dynamics during serial-dilution cycles using experimentally measurable quantities: Suppose a set of species in repeated serial-dilution cycles are observed to grow at growth rates $\{r_\alpha\}$ at time $t$ where the community has total biomass $\rhotot(t)$, to what extent can the set of data $\{r_\alpha,\rhotot\}$ recapitulate the existence of species and the dynamics of their abundances during the cycle? And how robust is the observed dynamics to perturbations in environmental factors and community composition? Most of the results derived in this study are centered around these questions.

One major finding is that community states cannot be taken for granted as ``niches'' - even when species ``take turns'' dominating growth in different community states, many species can go extinct. Instead, we find quantitative constraints on how fast or for how wide an interval the dominant species in each community state niche can grow. These constraints can be summarized as: (1) \emph{Tolerance to growth dominance:} increasing the growth rate of an individual species in its favored community state beyond a point does not impact coexistence. This effect contrasts starkly with steady-state coexistence, where the increased growth rate of one species can drive other species extinct. 
(2)  \emph{Community stability through staggered growth dominance:} Stability of a diverse community requires minimizing simultaneous fast growth of multiple species; that is, stability requires that species stagger their growth dominance across distinct community states.
(3) \emph{Increased growth dominance for late species:} species growing in late community states must grow faster or for larger biomass intervals than species in early states.

By showing that such niches can arise for a distinct mechanistic reason -- transitions between physiological states -- the Community State model makes distinct predictions about the relationship between physiology and ecology. Unlike in many other models, niches here are not created by a balance between microbes in exponential growth but originate through an interplay of switches between multiple growing and non-growing states. Our mutual invasibility criteria \eqref{eqn:IAB} offer a quantitative and intuitive understanding of the nature of these niches, factors that widen them, and the nature of competition between species in overlapping niches. The explicit dependence on the ratio of the growth rate of the invading species to that of the resident species in $I_{A,B}$ and $I_{B,A}$ make them a fitness-like measure for the current context (cyclic environments with multiple physiological states) where other fitness measures (e.g., growth rate difference) are not applicable. 

These effects persist in extensions of the model to many-species communities provided that the number of community states does not greatly exceed the growth dominance $p$, where $p$ is the ratio of the growth rate of the dominant species in a community state to the basal growth rate of subordinate species in that state. Trajectories of species abundances (Fig.~\ref{fig:off_diag}H) show that the dynamics in such systems no longer follow orderly succession dynamics but instead, show a seemingly-disordered array of growth curves that are nevertheless cyclic and hence maintain coexistence. To our knowledge, such disorderly, yet cyclic growth characteristics represent a new class of non-steady-state dynamics that has not been described in the ecology literature.  

The tendency in ecology, with an emphasis on steady states, has been to ``coarse-grain'' or ignore dynamics seen in real systems. Indeed, dynamics in some communities are merely complications (e.g., periodic perturbations about a stable steady state) that can be coarse-grained without any loss of understanding. In fact, in 1973, Stewart and Levin noted mathematically that two species could survive on a single ``seasonal resource''~\cite{stewart1973partitioning}. Their work has often been dismissed (including in their own paper~\cite{stewart1973partitioning}) as a mathematical observation relevant only for an assumed growth-affinity trade-off, narrow resource competition, and other idealizations. Our work argues that their simple mechanism of dynamic coexistence -- also explored in recent works~\cite{fridman2022fine,fink2022microbial,bloxham2022diauxic,bloxham2023biodiversity,wang2021complementary,burkart2022periodic} -- is \emph{more} relevant, not less, given the observed complex physiology and nonlinear growth dependencies in real microbial ecosystems. If physiological state changes turn out to be dominant drivers of dynamical niches, as seen in~\cite{amarnath2023stress,crocker2023genomic,goldford2018emergent}, dynamics cannot be ``averaged'' over but become the essential link between physiology and ecology.  


We regard an attractive feature of the top-down Community State model to be its direct quantitative connection to experimentally-accessible variables, as well as its avoidance of often inaccessible interaction parameters. Instantaneous growth rates of individual species can be obtained from transient changes in species abundances (via e.g., 16S sequence as proxy), and the total community biomass can be obtained by measuring total protein or total RNA as proxies, or simply by the optical density if the culture does not aggregate. The model studied here, therefore, provides a roadmap for the quantitative analysis of community-wide data to learn about community dynamics, going beyond taxonomic characterization, without invoking fitting parameters. This contrasts starkly with dynamical analysis based on commonly used bottom-up models which invariably involve a large number of unconstrained interaction parameters (e.g., the species interaction matrix in generalized Lotka-Volterra models, or the nutrient consumption matrix in Consumer-Resource models). Additionally, it emphasizes intra-cycle dynamics which has been largely neglected except for a few recent studies~\cite{bloxham2022diauxic,bloxham2023biodiversity,amarnath2023stress,crocker2023genomic}, and gives concrete predictions, e.g., on the growth rate and duration of early vs late species, that can be tested directly by data. In this sense, the Community State model is a phenomenological model that can be updated directly from data. 

Our approach shares common elements with other top-down approaches like the Stochastic Logistic Model~\cite{ grilli2020macroecological,zaoli2021macroecological} and recent data-driven models~\cite{Shahin2022-cx,ho2022competition} without explicit interspecies interactions. 
While these other models attribute growth rate fluctuations to external factors, our model focuses on endogenously-driven environmental change. Our model can be extended to incorporate external fluctuations that randomly perturb growth niches, either across hosts or across cycles, predicting various abundance distributions as in \cite{grilli2020macroecological}. However, a key distinction is that our approach imposes closure conditions on growth rate variations in repeated cycles needed for stable but dynamic coexistence.


The key idea in our work is the existence of global community states that can be sensed by microbes in that community. 
Our results suggest that it would be advantageous for organisms to use this information to adjust their behavior and grow in specific community states since such regulation would maximize their chance of survival in the community. For example, organisms occupying early phases of the cycle may benefit from limiting their own growth so as not to eliminate other species active later in the cycle, as late species could be important for the survival of all species in later phases of the cycle -- as is the case for acid-induced stress relief~\cite{amarnath2023stress}, the early blooming acid-producing sugar eater is rescued from death by the late-blooming acid consumer which removes the excreted acid and restores the environment. 

A more speculative aspect of the Community State model is that the sequence of community states can be parameterized by a one-dimensional eco-coordinate (as opposed to environmental factors or abundances of individual species). A further assumption that allowed for deriving quantitative coexistence criteria and relating them to empirical data is that community biomass can serve as this coordinate parameterizing the sequence of community states. We believe this hypothesis is biologically plausible: First, a number of key physiological parameters, e.g., pH, oxygen content, waste products, and iron availability, change with the accumulation of community biomass~\cite{okpokwasili2006microbial}, and the values of these parameters to cause transitions in the physiological states of individual organisms are known. Other physiological effects such as lag time and cell death might introduce limitations for our framework that require further study~\cite{manhart2018trade,bloxham2022diauxic,bloxham2023biodiversity}.
Second, it is known that several autoinducers are produced and sensed by a wide range of both gram-positive and gram-negative bacteria~\cite{xavier2003luxs,duan2003modulation,decho2010quorum}. In fact, AI-2 has been proposed to serve as a ``universal signal'' for inter-species communication~\cite{vendeville2005making,de2006let,galloway2011quorum}. Third, it is common for microbes to develop sensors to detect important features of their environment~\cite{janausch2002c4,eriksson2002low,o2016sensational,krulwich2011molecular}; as total community biomass is clearly an important dynamical variable that can be used to forecast the fate of the community (e.g., how close to the carrying capacity), it would not be surprising if organisms have evolved various proxy schemes to sense the total biomass. As bacteria feature multiple sensors and regulatory processes, they may detect various (and possibly distinct from other species) aspects of the global state of the community and integrate the available information through diverse regulatory mechanisms. Thus, community biomass may be viewed as a simplified description to summarize the effects of the different sensors. 


The defining feature of the Community State model is the ability of organisms in a community to sense common features of the community and their ability to modulate their own physiology in response to such community-wide signals. Indeed, the existence of a group of organisms that can sense and respond to common features in the environment may be taken as a key characteristic that defines a ``community''.


\section*{Acknowledgments}
This work was initiated at the Microbial ecology workshop that AM and TH participated in at the NSF-sponsored Kavli Institute of Theoretical Physics (NSF PHY-1748958.). We are grateful for helpful discussions with numerous colleagues during the course of this work, including Milena Chakraverti-Wuerthwein, Otto Cordero, Jacopo Grilli, Akshit Goyal, K.C. Huang, Seppe Kuehn, Pankaj Mehta, Ned Wingreen, and members of the Hwa lab. AVN and TH are supported by the Simons Foundation through the Principles of Microbial Ecosystems (PriME) collaboration (Grant no. 542387) and by the NSF (MCB-2029480). AM is supported by NIGMS of the NIH under award number R35GM151211 and the NSF through the Center for Living Systems (grant no.  2317138).

\section*{Methods}
All numerical results in Figs.~1 and 2 were obtained by simulations performed in Matlab (code available at https://github.com/avaneeshnarla/dynamic-metabolic). All other results were obtained using simulations performed in Python 3 using forward Euler integration. For the results in Fig.~4, the integration was performed in time with a growth period of 20 units and a time resolution of 0.001 units. For Fig.~5, the integration was performed in the normalized biomass coordinate ($\eta$) ranging from 0 to 1, with a resolution of 0.001 units distributed evenly between the niches (such that each niche required the same number of forward integration steps). The initial population abundances in all cases were random fractions of $\delta\cdot\cmax$, a dilution of the maximum biomass attainable by the population as per our model. Random numbers were drawn from a uniform distribution using Python 3's Random package.

\printbibliography

\end{document}


\author[1]{Avaneesh V. Narla}
\author[1]{Terence Hwa} 
\author[2]{Arvind Murugan}

\affil[1]{Department of Physics, University of California, San Diego}
\affil[2]{Department of Physics, University of Chicago}

\maketitle

\vspace{-18pt}
\tableofcontents




\section{Supplementary Figures}
\begin{figure}[!ht]
 \includegraphics[width=\columnwidth]{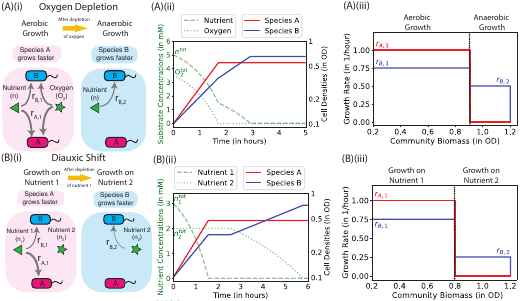}
 \caption[Growth transitions arising in oxygen depletion and diauxic shift]{\textbf{Growth transitions arising in oxygen depletion and diauxic shift.} Illustrations are
made for two species, A (red) and B (blue), whose growth rates, $r_A$ and $r_B$, respectively, vary due to
changes in the concentrations of nutrients or toxins in the common media due to a variety of
interactions. Growth curves are shown as red and blue solid lines in the middle column, with nutrient/toxin
concentrations shown as green dashed or dotted lines. Growth rates are plotted against the
accumulated total biomass density in the right column. \textbf{A.} Species A grows faster than B aerobically.
However, after the depletion of oxygen, only B continues to grow anaerobically, although at a slower
rate. \textbf{B.} Species A grows faster than B on nutrient 1. Species A stops growing after the depletion of nutrient 1 but B continues to grow on nutrient 2 after a classic diauxic shift, reflected by a lag in the growth
curve (middle column of panel B). Note that when plotting the growth rate against the total biomass
accumulated in the community, the lag does not show up as there is no biomass accumulation
during this period. More such examples are provided in Fig.~\ref{map2}}
 \label{map}
\end{figure}

\begin{figure}[H]
    \centering
    \includegraphics[width=\columnwidth]{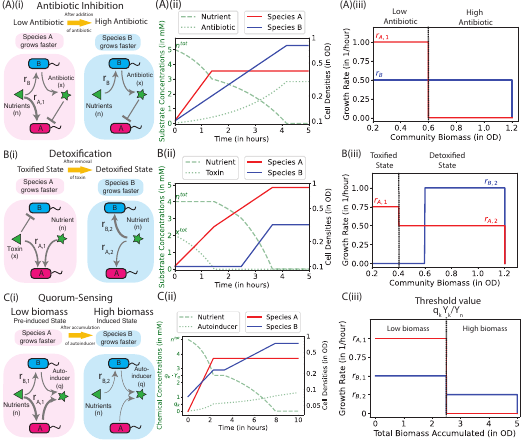}
    \caption[Growth transitions arising in antibiotic inhibition, detoxification, and quorum-sensing]{\textbf{Growth transitions arising in antibiotic inhibitions, detoxification, and quorum-sensing.} Illustrations are
made for two species, A (red) and B (blue), whose growth rates, $r_A$ and $r_B$, respectively, vary due to
changes in the concentrations of nutrients or toxins in the common media due to a variety of
interactions. Growth curves are shown as red and blue lines in the middle column, with nutrient/toxin
concentrations shown as the green dashed or dotted lines. Growth rates are plotted against the
accumulated total biomass in the right column. \textbf{A.} Species A grows faster than Species B, but Species B excretes an antibiotic which inhibits the
growth of Species A when accumulated to a sufficient level. B itself is not sensitive to the antibiotic. \textbf{B.} A toxin
in the medium inhibits the growth of Species B, but is consumed by Species A. After toxin removal, A
continues to grow (at a moderately reduced rate) while B can grow at a fast rate. \textbf{C.} Both species secrete autoinducers. After the autoinducer concentration reaches a threshold value, B grows faster than A.}
    \label{map2}
\end{figure}


\begin{figure}[H]
    \centering
    \includegraphics[width=0.8\columnwidth]{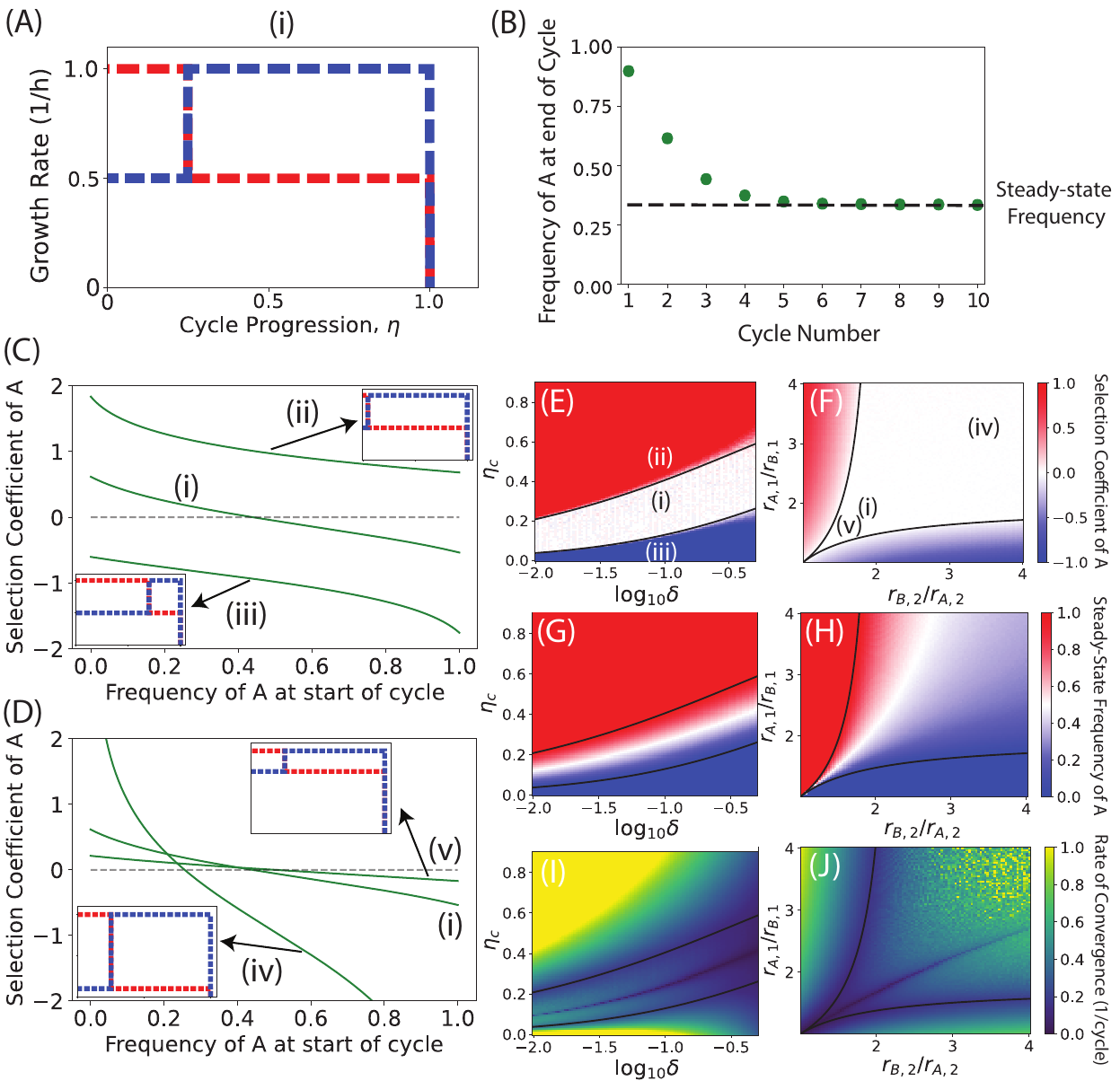}
    \caption{\textbf{Partitioning of growth rates reveals quality of coexistence.} Features from growth rate dependences on $\eta$ can be understood by considering perturbations to a piece-wise linear growth rate dependence of two species as shown in (A) and hereby denoted by (i). For all parameters, the coculture eventually arrives at a steady-cycle frequency at the end of the cycle that does not change in subsequent cycles as shown in (B). All results indicated below are for the steady cycle. (C,D) The variation of selection coefficient of A with initial frequency of A for different growth rate dependences (shown in insets and labeled). If the selection coefficient is not 0 for any initial frequency, the two species will not coexist. The effect of the perturbations can be understood more systematically in phase plots where either the environment (in panels E,G,I) or the physiology (in panels F,H,J) of the two species is varied while holding all other parameters constant.}
    \label{fig:fig_selection}
\end{figure}

\begin{figure}[H]
    \centering
    \includegraphics[width=\columnwidth]{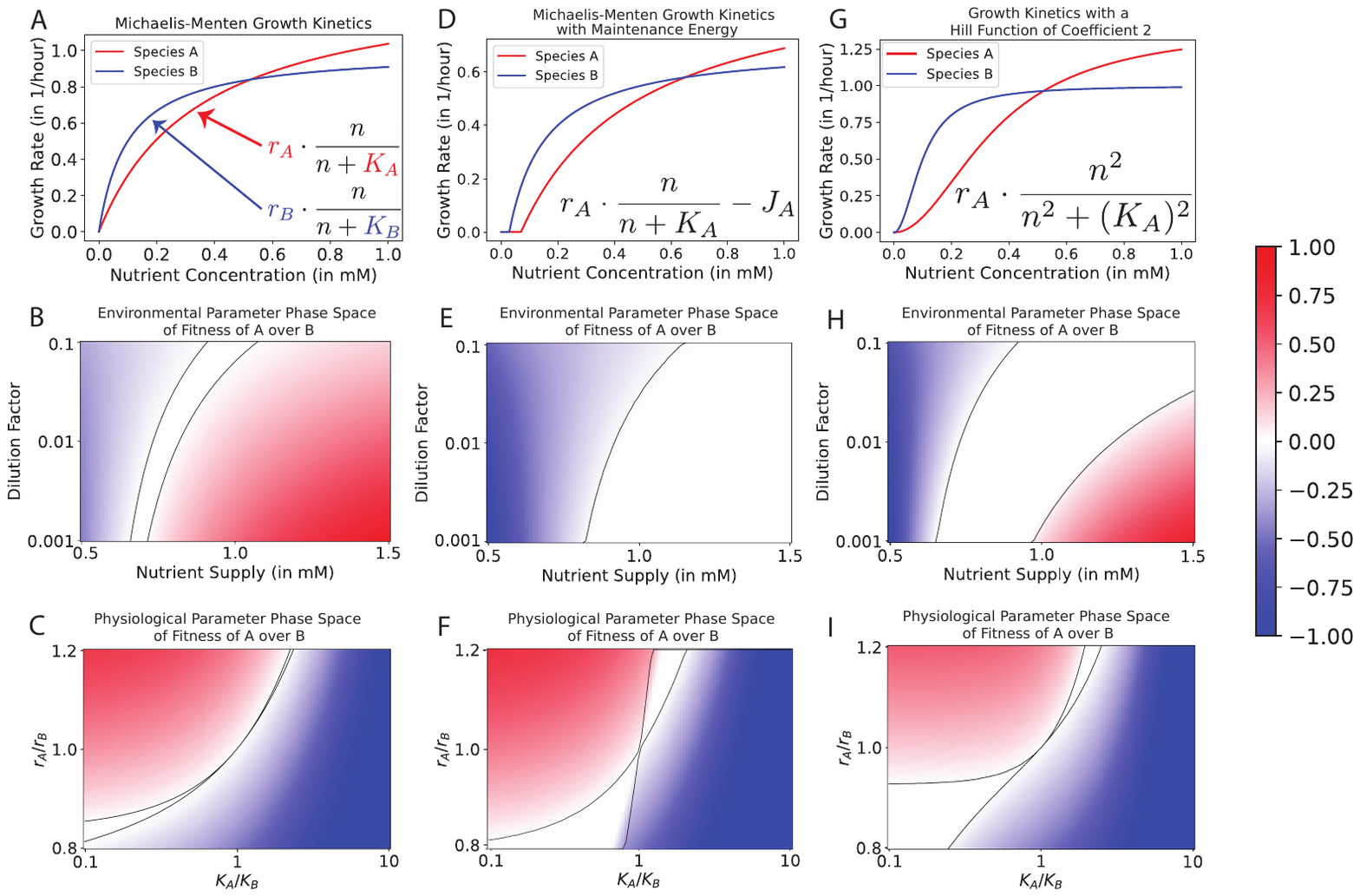}
    \caption{\textbf{Biologically-motivated modifications to the Monod growth relation enlarge the coexistence region of phase space.} \textbf{A}. Plot showing the standard Monod Growth relation for the dependence of the growth rate on the nutrient concentration of the medium. Two species with different constants describing the growth rates ($r_i$ and $K_i$) are shown and labeled as Species A and B. \textbf{B, E, and H}. Plot showing the fitness of species A over species B over one cycle after 100 cycles (common colorbar shown to the right) for different values of $r_A/r_B$ and $K_A/K_B$ for the respective growth functions of each row. The black lines indicate the boundaries of the analytically determined phase boundaries of coexistence. \textbf{B}. Plot showing the fitness of species A over species B over one cycle after 100 cycles (common colorbar shown to the right) for different values of environmental parameters (the nutrient supplied at the beginning of each growth cycle and the dilution factor). The black lines indicate the boundaries of the analytically determined phase boundaries of coexistence. \textbf{C}. Plot showing the fitness of species A over species B over one cycle after 100 cycles (common colorbar shown to the right) for different values of physiological parameters of species A, shown as ratios $r_A/r_B$ of the two growth rates and $K_A/K_B$ of the two saturation constant. The black lines indicate the boundaries of the analytically determined phase boundaries of coexistence.}
    \label{monod}
    \end{figure}
    \begin{figure}[H]
    \centering
    \includegraphics[width=\columnwidth]{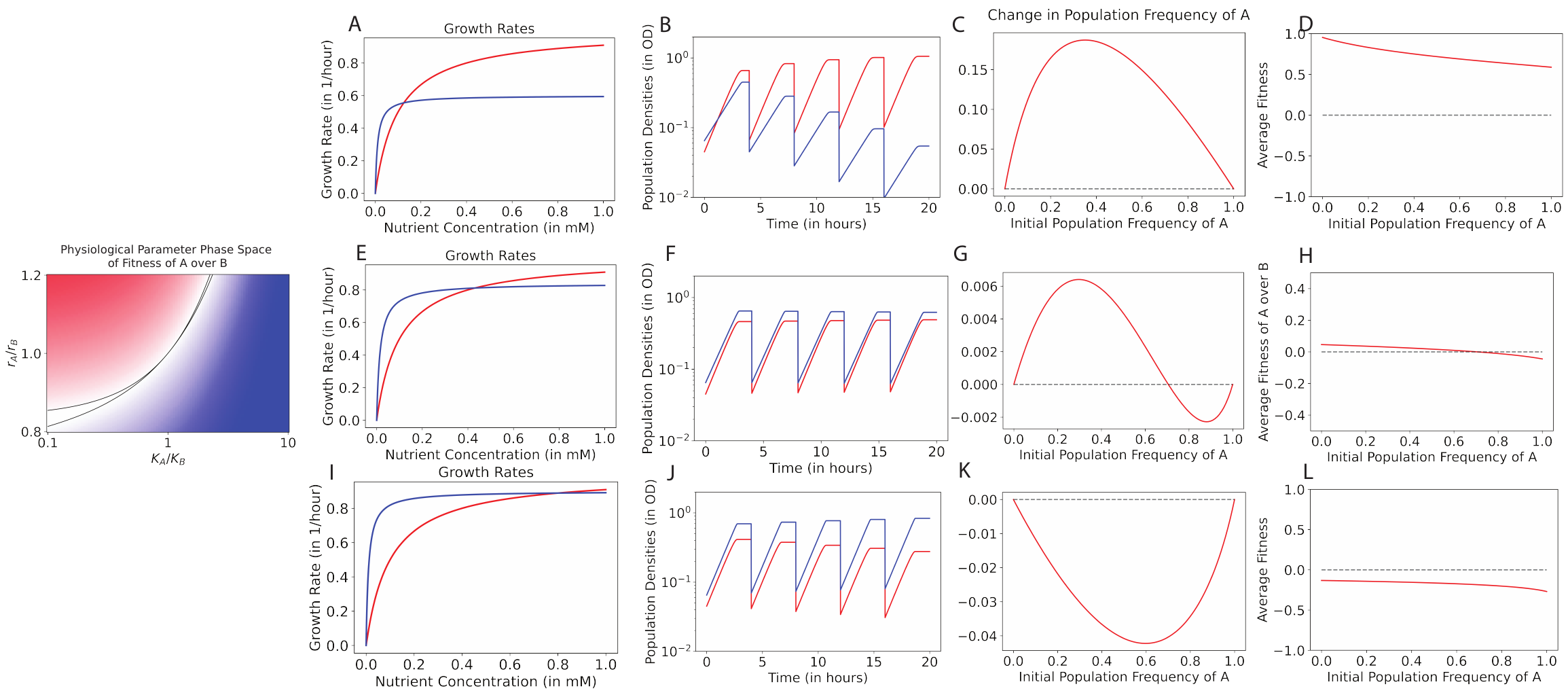}
    \caption{\textbf{Assembly of complex communities can be understood by the mutual invasibility criterion for coexistence.} \textbf{A.} The growth functions of two species in competition (either of the species can be replaced by a community with the growth rate indicating the growth rate of the entire community, averaged by the frequency of each member of the community). This is determined entirely by the physiological parameters of the two strains. \textbf{B.} The ratios of the two growth rates indicate the differential fitness within the cycle. \textbf{C.} The differential fitness needs to be weighted by a resource consumption kernel, given by $\omega(s)$ which is a function of the environmental variables, $s_0$ and $\delta$. If the integral of $r_A(s)/r_B(s)$ curve is $\geq 1$, Species A can invade the monoculture of B, and similarly for A. \textbf{D.} Iterative flow maps for both Species A and species showing that if both monocultures are invadible, then one fixed point must exist. The invasibility criteria allows one to infer the existence of a coexistence fixed point. If either monoculture is invadable, the trivial fixed point of that monoculture is unstable. As shown in the text, only one non-trivial fixed point can exist.} 
    \label{fig:fig5}
\end{figure}

\begin{figure*}
    \centering
    \includegraphics[width=\columnwidth]{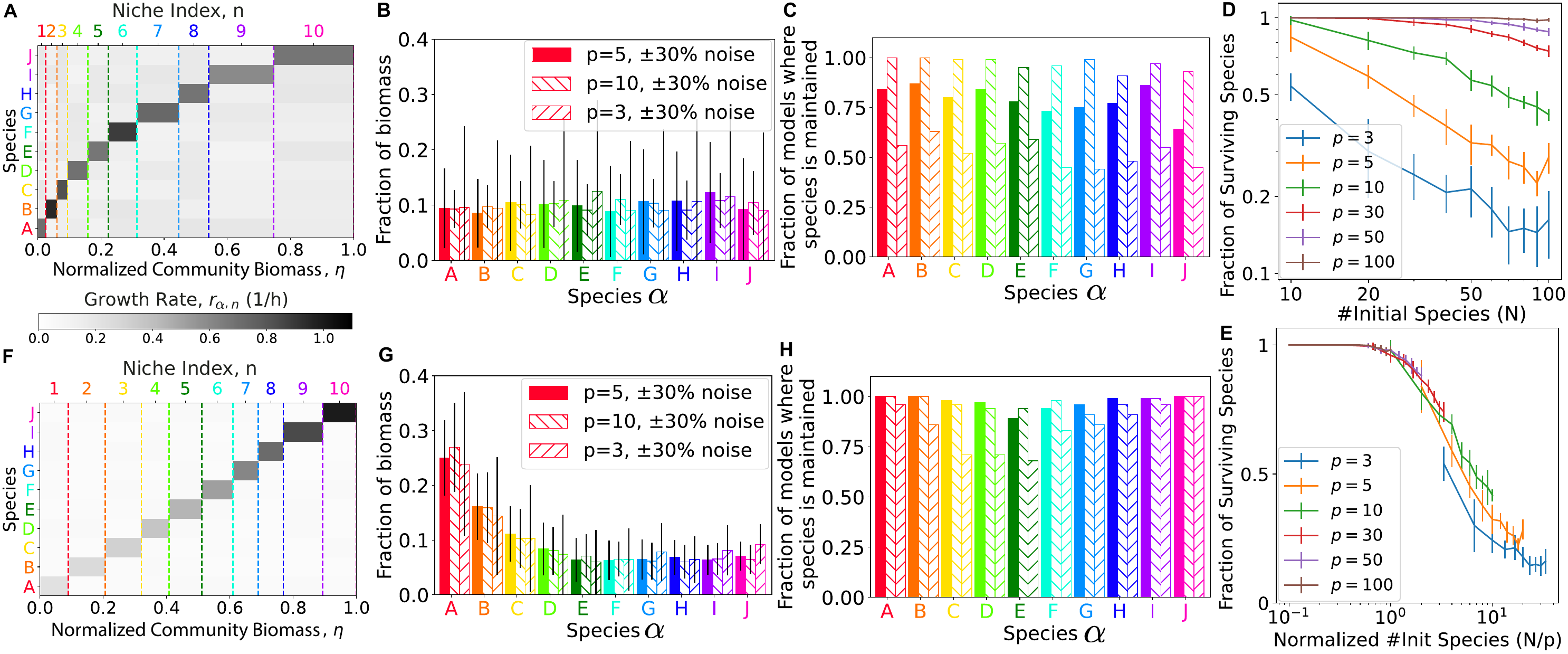}
    \caption{
    \textbf{Impact of disorder and relative changes in growth rate across niches on coexistence}
    (A) Model with exponential niche width, with the value of growth rate for each diagonal entry (i.e., during the preferred niche) assigned randomly within a range of $r_+$, each off-diagonal entry (non-preferred niches) assigned randomly within a range of $r_-$. (B) The abundance of each species at the end of the stable cycle, obtained for the model parameters indicated, with $p\equiv r_+/r_-$ being the growth preference. The black lines indicate the standard deviation. (C) The fraction of models where each species is maintained, for the same set of parameters as those indicated in the legend of panel B.
    (D, E) Abundance of the surviving species for different number of niches $N$ and different growth preferences. Panels F, G, H are the same as panels A, B, C, except that niche widths are fixed to a constant, but the growth preference $r_+/r_-$ is varied using \eqref{r-eta}}
    \label{fig:many_noise}
\end{figure*}

\begin{figure}
    \centering
    \includegraphics[scale=0.6]{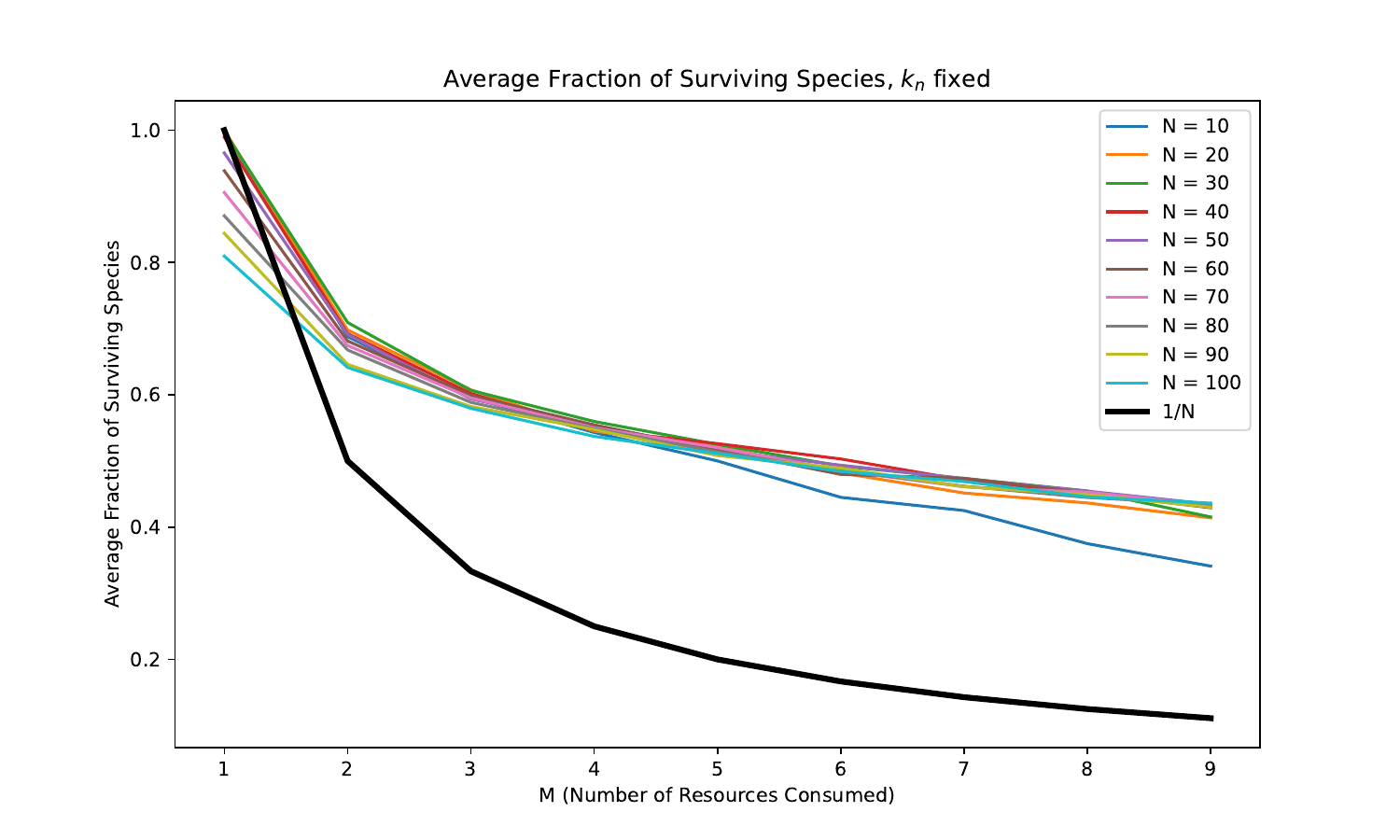}
    \caption{Average Fraction of Surviving Species with $K_n=M$: The plot shows the average fraction of surviving species in a chemostat model where the number of resources consumed per species ($K_n$) is held constant. The survival fractions are plotted against varying numbers of resources consumed per species ($M$). Each line represents a different total number of species/resources ($N$). The solid black line is $1/M$.}
    \label{fig:CRM-kn}
\end{figure}

\begin{figure}
    \centering
    \includegraphics[scale=0.6]{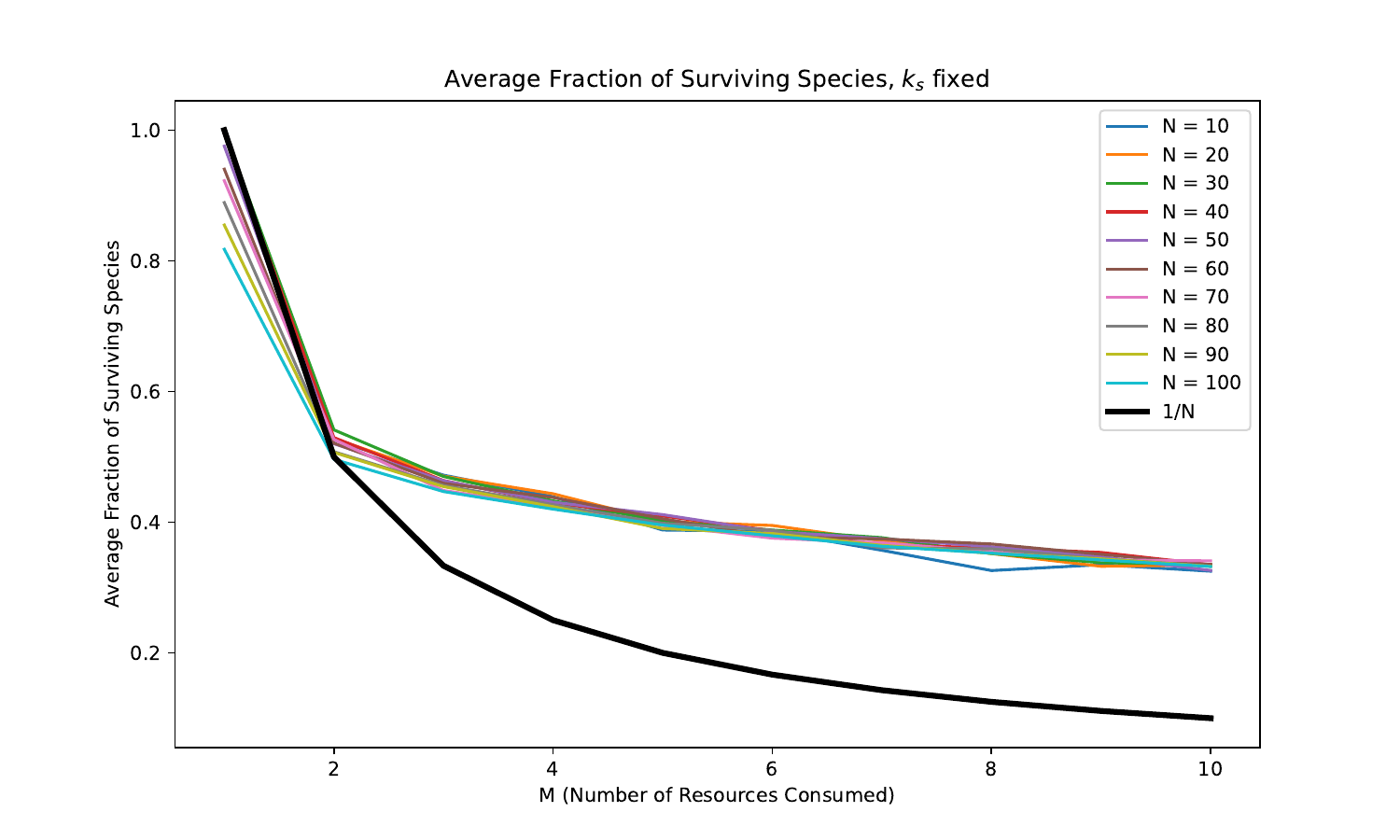}
    \caption{Average Fraction of Surviving Species with $K_s=M$: The plot shows the survival fractions of species under the condition where the number of species consuming each resource ($K_s$) is constant. The plot shows how species survival varies with the change in the number of resources consumed per species ($M$). Different lines correspond to different total numbers of species/resources ($N$). The solid black line is $1/M$.}
    \label{fig:CRM-ks}
\end{figure}


\newpage
\section{Supplementary Text}
\subsection{Context on the Competitive Exclusion Principle}
 The Competitive Exclusion Principle (CEP) states that complete competitors cannot coexist~\cite{hardin1960competitive}. The statement originates with Volterra who used a mathematical model to demonstrate that two species whose growth is limited by the same resource cannot coexist indefinitely~\cite{volterra1928variations}. This idea was further explored, developed, and disseminated by Lotka~\cite{lotka1978growth}, Gause~\cite{gause2019struggle,gause1934experimental}, and Hutchinson~\cite{hutchinson1957population}. Subsequent theoretical work by MacArthur, Levins, and others~\cite{macarthur1964competition,macarthur2016theory,levins1968evolution,rescigno1965competitive} extended the CEP to state that, in general, there can be no more species than resources.\par
The CEP is also closely tied to the notion of an ecological niche~\cite{grinnell1904origin,grinnell1917niche} and is alternatively stated as ``No two species can indefinitely continue to occupy the same ecological nich''~\cite{slobodkin1961growth}. Levin showed that two species cannot occupy a niche (defined as the hypervolume where the dimensions are environmental conditions and resources, following Hutchinson's definition~\cite{hutchinson1957population}) unless their limiting factors (for example, nutrients) differ and are independent~\cite{levin1970community}. This echoes MacArthur when he says that the proper statement of the CEP is that ``species divide up the resources of a community in such a way that each species is limited by a different factor.''~\cite{macarthur1958population}. We must note that this holds for both biotic and abiotic factors~\cite{volterra1928variations,macarthur1964competition,levins1968evolution,levin1970community}.\par
All the attempts described above contained the assumption that the specific growth rates of the competing species are linear functions of resource or factor densities~\cite{armstrong1980competitive}. Further, most attempts (with the exception of Levin~\cite{levin1970community}) only considered coexistence at fixed densities. In fact, it can be shown that coexistence at fixed densities is limited by the number of resources the species can grow on~\cite{armstrong1980competitive}, regardless of the form of the growth rates. When both these constraints are simultaneously relaxed, many species can coexist on a few biotic resources~\cite{koch1974competitive,mcgehee1977some,armstrong1980competitive}. A key result was provided by Levins who showed that the number of effective resources is the number of original resources plus the number of distinct non-linearities in the system~\cite{levins1979coexistence}.\par
Constructive examples rely on the periodic solutions of the Lotka-Volterra model~\cite{volterra1928variations} and different points of saturation in the non-linearities of the specific growth rates (we note that nonlinear saturating functional responses are more biologically accurate than linear growth rates~\cite{holling1959components}).  For the case of abiotic resources, Smale showed that that the ordinary differential equation commonly used to describe competing species are compatible with any dynamical behavior provided the number of species is greater than three~\cite{smale_1976}. Following this, systems have been constructed that have periodic orbits~\cite{armstrong1976coexistence} with more than three species. Even chaotic behavior has been observed for the case of essential nutrients~\cite{huisman1999biodiversity}. However, such constructed examples require a large number of species and/or nutrients, need to be carefully constructed, and there is no evidence that they occur naturally. Concurrently, in the 1970s, the importance of non-equilibrium interactions of competing populations in establishing species diversity was emphasized in several investigations~\cite{huston1979general} (See Chapter 15 of \cite{mittelbach2019community}).\par

\subsection{Mapping between different mechanistic models for step-wise growth}
\label{map_text}
\subsubsection{Base Model}
We first describe the base model for the growth step of the growth-dilution cycle, and below we will map different consumer-resource models to the base model mathematically.

We take $\rho_A$ to be the population of Species A, $\rho_B$ to be the population of Species B, and $\eta$ to be the eco-coordinate.
\begin{align}
    \dot{\rho_A}&=r_A(\eta)\cdot \rho_A,\label{base1}\\
    \intertext{where $r_A(\eta)=r_{A,1}$ when $\eta<\eta_k$ and $r_A(s)=r_{A,2}$ when $\eta>\eta_k$.}
    \dot{\rho_B}&=r_B(\eta)\cdot \rho_B,\label{base2}\\
    \intertext{where $r_B(\eta)=r_{B,1}$ when $\eta<\eta_k$ and $r_B(s)=r_{B,2}$ when $\eta>\eta_k$.}
    \dot{\eta}&=r_A(\eta)\cdot \rho_A+r_B(\eta)\cdot \rho_B.\label{base3}
\end{align}
$\eta$ starts at 0 and ends at 1.
\subsubsection{Step-Wise Growth}
We take $N_A$ to be the population of Species A, $N_B$ to be the population of Species B, and $s$ to be the nutrient. The consumer resource model is given by
\begin{align}
    \dot{N_A}&=r_A(s)N_A,\\
    \intertext{where $r_A(s)=r_{A,1}$ when $s>s_k$ and $r_A(s)=r_{A,2}$ when $s<s_k$.}
    \dot{N_B}&=r_B(s)N_B,\\
    \intertext{where $r_B(s)=r_{B,1}$ when $s>s_k$ and $r_B(s)=r_{B,2}$ when $s<s_k$.}
    \dot{s}&=-r_A(s)N_A/Y_A-r_B(s)N_B/Y_B.
\end{align}
The environmental variables are given by $s_0$, the total amount of nutrient supplied at first, and $\delta$, the dilution fold.\\
We redefine the variables as follows:
\begin{align}
\eta&\equiv 1-\frac{s}{s_0}\\
\eta_k&\equiv 1-\frac{s_k}{s_0}\\
\rho_A&\equiv \frac{N_A}{s_0\cdot Y_A}\\
\rho_B&\equiv \frac{N_B}{s_0\cdot Y_B}
\end{align}
This creates an exact equivalence to the base model.
\subsubsection{Diauxie}
We take $N_A$ to be the population of Species A, $N_B$ to be the population of Species B, $s_1$ to be the concentration of the first nutrient consumed and $s_2$ to be the concentration of the second nutrient consumed. The consumer resource model is given by
\begin{align}
    \dot{N_A}&=r_{A,1}(s_1)N_A+r_{A,2}(s_1,s_2)N_A,\\
    \intertext{where $r_{A,1}(s_1)=r_{A,1}\cdot \Theta(s_1)$ and $r_{A,1}(s_1,s_2)=r_{A,2}\cdot \Theta(-s_1)\cdot \Theta(s_2)$. $\Theta$ is the Heaviside step-function}
    \dot{N_B}&=r_{B,1}(s_1)N_B+r_{B,2}(s_1,s_2)N_B,\\
    \intertext{where $r_{B,1}(s_1)=r_{B,1}\cdot \Theta(s_1)$ and $r_{B,1}(s_1,s_2)=r_{B,2}\cdot \Theta(-s_1)\cdot \Theta(s_2)$.}
    \dot{s_1}&=-r_{A,1}(s_1)\frac{N_A}{Y_A\cdot Y_1}-r_{B,1}(s_1,s_2)\frac{N_B}{Y_B\cdot Y_1}.\\
    \dot{s_2}&=-r_{A,2}(s_2)\frac{N_A}{Y_A\cdot Y_2}-r_{B,2}(s_1,s_2)\frac{N_B}{Y_B\cdot Y_2}.
\end{align}
The environmental variables are given by $s_0^1$ and $s_0^2$, the total amount of nutrients supplied at first, and $\delta$, the dilution fold.\\
We redefine the variables as follows:
\begin{align}
\eta&\equiv \begin{cases}1-\frac{Y_1\cdot s_1+Y_2\cdot s_0^2}{(Y_1\cdot s_0^1+Y_2\cdot s_0^2)},s_1>0\\1-\frac{Y_2\cdot s_2}{(Y_1\cdot s_0^1+Y_2\cdot s_0^2)},s_1=0\end{cases}\\
\eta_k&\equiv 1-\frac{Y_1\cdot s_0^1}{(Y_1\cdot s_0^1+Y_2\cdot s_0^2)}\\
\rho_A&\equiv \frac{N_A}{Y_A\cdot{(Y_1\cdot s_0^1+Y_2\cdot s_0^2)}}\\
\rho_B&\equiv \frac{N_B}{Y_B\cdot{(Y_1\cdot s_0^1+Y_2\cdot s_0^2)}}
\end{align}
This gives us that:
\begin{align}
    \dot{\rho_A}&=r_A(\eta)\rho_A,\\
    \intertext{where $r_A(\eta)=r_{A,1}\cdot \Theta(\eta_k-\eta)$ and $r_A(\eta)=r_{A,2}\cdot \Theta(\eta-\eta_k)\cdot \Theta(1-\eta)$.}
    \dot{\rho_B}&=r_B(\eta)\rho_B,\\
    \intertext{where $r_B(\eta)=r_{B,1}\cdot \Theta(\eta_k-\eta)$ and $r_B(\eta)=r_{B,2}\cdot \Theta(\eta-\eta_k)\cdot \Theta(1-\eta)$.}
    \dot{\eta}&=r_A(\eta)\rho_A+r_B(\eta)\rho_B.
\end{align}
This creates an exact equivalence to the base model.
\subsubsection{Oxygen Depletion}
We take $N_A$ to be the population of Species A, $N_B$ to be the population of Species B, $s$ to be the concentration of the nutrient, and $x$ to be the concentration of oxygen. The consumer resource model is given by
\begin{align}
    \dot{N_A}&=r_A(s,x)N_A,\\
    \intertext{where $r_A(s,x)=r_{A,1}\cdot \Theta(x)+r_{A,2}\cdot \Theta(-x)\cdot \Theta(s)$.}
    \dot{N_B}&=r_B(s,x)N_B,\\
    \intertext{where $r_B(s,x)=r_{B,1}\cdot \Theta(x)+r_{B,2}\cdot \Theta(-x)\cdot \Theta(s)$.}
    \dot{s}&=-r_A(s,x)N_A/Y_A-r_B(s,x)N_B/Y_B.\\
    \dot{x}&=(-r_{A,1}\cdot \frac{N_A}{Y_A\cdot Y_x}-r_{B,1}\cdot \frac{N_B}{Y_B\cdot Y_x})\Theta(x).
\end{align}
The environmental variables are given by $s_0$ and $x_0$, the total amount of nutrient and oxygen supplied at first, and $\delta$, the dilution fold. For oxygen to deplete first, we need $x_0<s_0\cdot Y_x$\\
We redefine the variables as follows:
\begin{align}
\eta&\equiv 1-\frac{s}{s_0}\\
\eta_k&\equiv 1-\frac{x_0}{s_0\cdot Y_x}\\
\rho_A&\equiv \frac{N_A}{s_0\cdot Y_A}\\
\rho_B&\equiv \frac{N_B}{s_0\cdot Y_B}
\end{align}
This creates an exact equivalence to the base model.
\subsubsection{Quorum Sensing}
We take $N_A$ to be the population of Species A, $N_B$ to be the population of Species B, $s$ to be the concentration of the nutrient, and $q$ to be the concentration of autoinducer. The consumer resource model is given by
\begin{align}
    \dot{N_A}&=r_A(s,x)N_A,\\
    \intertext{where $r_A(s,x)=r_{A,1}$ when $q<q_k$, $r_{A,2}$ when $q>q_k$, and 0 when $s=0$.}
    \dot{N_B}&=r_B(s,x)N_B,\\
    \intertext{where $r_B(s,x)=r_{B,1}$ when $q<q_k$, $r_{B,2}$ when $q>q_k$, and 0 when $s=0$.}
    \dot{s}&=-r_A(s,x)N_A/Y_A-r_B(s,x)N_B/Y_B.\\
    \dot{q}&=r_{A,1}\cdot \frac{N_A}{Y_A\cdot Y_q}+r_{B,1}\cdot \frac{N_B}{Y_B\cdot Y_q}.
\end{align}
The environmental variables are given by $s_0$, the total amount of nutrient supplied at first, and $\delta$, the dilution fold. For the autoinducer to accumulate first, we need $q_k<s_0\cdot Y_q$\\
We redefine the variables as follows:
\begin{align}
\eta&\equiv 1-\frac{s}{s_0}\\
\eta_k&\equiv 1-\frac{q_k}{s_0\cdot Y_q}\\
\rho_A&\equiv \frac{N_A}{s_0\cdot Y_A}\\
\rho_B&\equiv \frac{N_B}{s_0\cdot Y_B}
\end{align}
This creates an exact equivalence to the base model.
\subsubsection{Acid Stress}
We take $N_A$ to be the population of Species A, $N_B$ to be the population of Species B, $s$ to be the concentration of the nutrient, $x$ to be the concentration of acid, and $p$ to be the concentration of pyruvate. The consumer resource model is given by
\begin{align}
    \dot{N_A}&=r_A(s,x,p)\cdot N_A,\\
    \intertext{where $r_A(s,x)=r_{A,1}$ when $x<x_c$, $r_{A,2}$ when $x>x_c$, and 0 when $s=0$.}
    \dot{N_B}&=r_B(s,x,p)\cdot N_B,\\
    \intertext{where $r_B(s,x)=r_{B,1}$ when $x<x_c$, $r_{B,2}$ when $x>x_c$, and 0 when $s=0$.}
    \dot{s}&=-r_A\cdot\frac{N_A}{Y_A\cdot Y_s}.\\
    \dot{x}&=r_A\cdot \frac{N_A}{Y_A\cdot Y_x}\cdot f_x-r_B\cdot \frac{N_B}{Y_B\cdot Y_x}.\\
    \dot{p}&=r_A\cdot \frac{N_A}{Y_A\cdot Y_p}\cdot f_p-r_B\cdot \frac{N_B}{Y_B\cdot Y_p}.
\end{align}
such that $f_x+f_p<1$ and the fractions of the nutrient consumed that Species A converts to acid and pyruvate.
The environmental variables are given by $s_0$, the total amount of nutrient supplied at first, and $\delta$, the dilution fold.
We redefine the variables as follows:
\begin{align}
\eta&\equiv 1-\frac{Y_s\cdot s+Y_x\cdot x+Y_p\cdot p}{Y_s\cdot s_0}\\
\rho_A&\equiv \frac{N_A}{s_0\cdot Y_A}\\
\rho_B&\equiv \frac{N_B}{s_0\cdot Y_B}
\end{align}
All that remains to resolve is the question of $\eta_k$ since in the actual system, it is only given by the acetate concentration, $x_c$. However, we can take $x_c$ to be a function of $\eta$ if $s(t)\propto x(t)$:
\begin{align}
\eta(x=x_c)&\equiv 1-\frac{Y_s\cdot s(x_c)+Y_x\cdot x_c}{Y_s\cdot s_0}
\end{align}
In reality,
\begin{align}
    s(t)&=N_A\cdot(\exp(r_{A,1}\cdot t)-1)\cdot \frac1{Y_A\cdot Y_s}\\
    x(t)&=N_A\cdot(\exp(r_{A,1}\cdot t)-1)\cdot \frac1{Y_A\cdot Y_x}-N_B\cdot(\exp(r_{B,1}\cdot t)-1)\cdot \frac1{Y_B\cdot Y_x}\\
    &=s(t)\cdot \frac{Y_s}{Y_x}-\frac{N_B}{Y_B\cdot Y_x}\cdot(\exp(r_{B,1}\cdot t)-1)\\
    &= s(t)\cdot \frac{Y_s}{Y_x}-\frac{N_B}{Y_B\cdot Y_x}\cdot\left({\frac{Y_A\cdot Y_s\cdot s(t)}{N_A}}\right)^{r_{B,1}/r_{A,1}}
\end{align}
Thus, $s(t)\propto x(t)$ if $r_{B,1}\ll r_{A,1}$ (or $N_B\ll N_A$). This creates an exact equivalence to the base model.
\subsection{Rescaling}
\label{rescale}
Here, we discuss the rescaling of the time spent in each phase/niche and the transition points in the piece-wise linear models discussed in the paper. The rescaling of the transition points amounts to a redefinition of the units with $\cmax$ declared to be 1, with an appropriate normalization of all transition points by $\cmax$. For the many species model, we additionally define a parameter, $\eta \equiv (\rhotot/\cmax-\delta)/(1-\delta)$, for simplification of the results such that the cycle starts at $\eta=0$ and ends at $\eta=1$.
For rescaling the time spent in each phase, we first note that in the limit of large time between dilution events, there is no activity once the system reaches the total biomass density of the system reaches $\cmax$. Thus, we may ignore the timescales associated with dilution. Next, we consider the dynamics within niche/phase, $n$:
\begin{equation}
    \frac{d}{dt}\rho^{(j)}_\alpha=r_{\alpha,n}\cdot \rho^{(j)}_\alpha(t) \qquad {\rm for }\ \eta_{n-1}\le \eta \le \eta_{n-1}.
\end{equation}
The dynamics in this interval is essentially independent of every other phase as the system is autonomous in time. Thus we may define, for example, the dimensionless time, $\tau_n\equiv t\cdot\langle\alpha r_{\alpha,n}\rangle_{\alpha\neq\alpha(n)}$. Hence, rescaling $t$ by $\tau_n$, we obtain that the dynamics of $\rho^{(j)}_\alpha$ in niche $n$ given by
\begin{equation}
    \frac{d}{d\tau_n}\rho^{(j)}_\alpha=\frac{r_{\alpha,n}}{\langle\alpha r_{\alpha,n}\rangle_{\alpha\neq\alpha(n)}}\cdot \rho^{(j)}_\alpha(t) \qquad {\rm for }\ \eta_{n-1}\le \eta \le \eta_{n-1}.
\end{equation}
This allows us to define $p_n\equiv \frac{r_{\alpha,n}}{\langle\alpha r_{\alpha,n}\rangle_{\alpha\neq\alpha(n)}}$. Thus, in the cases without noise described in this manuscript, each niche is effectively described by one number: the domination of the fast-growing species in niche $n$ over the slow-growing species.
\subsection{Mathematical Results for the Simple Toy Model}
Though we have shown that two species may coexist in growth-dilution cycles, understanding the dynamics of the co-culture system mathematically can be significantly difficult because of the non-linear growth rates involved. To simplify the system, we consider a piece-wise linear approximation of the system (as shown in Fig.~3A). In this Toy Model, we take Species A to grow at a constant rate $r_{A,1}$ and Species B at $r_{B,1}$ while biomass values are above a threshold $s_k$ (we call this the first phase and call the time taken to complete it $\tau_1$). And in the 2nd phase (biomass from $0$ to $\eta_k$, taking time $\tau_2$), we take Species A to grow at $r_{A,2}$ and Species B at $r_{B,2}$ (this toy model is described in Fig.~3A). When biomass reaches $\cmax$, we assume that both species stop growing.\par
If two species coexist in such a system, it implies that the net growth rate is the same over the steady-state cycle, i.e.,
\begin{align}
    r_{A,1} \tau_1+r_{A,2} \tau_2=r_{B,1} \tau_1+r_{B,2} \tau_2=-\log\delta.\label{r_tau}
\end{align}
\label{steady_comp}
We call the time taken to complete the first phase, $\tau_1$, and the time taken to complete the second phase, $\tau_2$. If two species coexist in such a system, it implies that the net growth rate is the same over the steady-state cycle, i.e.,
\begin{align}
 r_{A,1} \tau_1+r_{A,2} \tau_2=r_{B,1} \tau_1+r_{B,2} \tau_2=-\log\delta.\label{r_tau2}
\end{align}
For our purposes, we assume that the time between dilution steps, $T$, is longer than $\tau_1+\tau_2$.
We note that as long as there is a trade-off in the growth rates (such that neither species is growing faster than the other species at all biomass values), there exists a positive $\tau_1$ and $\tau_2$ that can solve \eqref{r_tau}. However, \eqref{r_tau} is also coupled with a set of equations describing the biomass increase that feature the populations of both species, $\rho_A(0)$ and $\rho_B(0)$, at the beginning of the cycle:
\begin{align} 
\thresh-\delta\cdot\cmax &= (e^{r_{A,1} \tau_1}-1) \rho_A(0) + (e^{r_{B,1} \tau_1} - 1 )\rho_B(0), \label{res_1}\\
\cmax-\thresh &= (e^{r_{A,2} \tau_2}-1)e^{r_{A,1} \tau_1} \rho_A(0) + (e^{r_{B,2} \tau_2} - 1 )e^{r_{B,1} \tau_1} \rho_B(0). \label{res_2}
\end{align}
We obtain the equations above as the total amount of biomass produced by both species must be $\thresh-\delta\cdot\cmax$ in the first phase (as the co-culture starts with biomass of $\delta\cdot\cmax$ after dilution by a factor of $\delta$ from the maximal biomass value of $\rhotot$, and $\cmax-\thresh$ in the second phase.
The solution to \eqref{res_1}-\eqref{res_2}, however, may not yield a viable positive solution for $\rho_A(0)$ and $\rho_B(0)$. For viable solutions, we require that the two species must necessarily engage in resource sharing in the steady-state cycle described by $\tau_1$ and $\tau_2$ that solve \eqref{r_tau}, as otherwise they would not be able to accumulate all of the biomass by themselves. If any resources remain unconsumed in a monoculture (say of Species A), a very small population of Species B can consume the remaining resources and grow more than the dilution fold. Eventually, the very small population grows larger and the factor of growth decreases until both species grow at the dilution fold.\par
Thus, to get $\rho_A(0)>0$, we must have (assuming $r_{A,1}>r_{B,1}>0,\ r_{B,2}>r_{A,2}>0$):
\begin{equation}
 \rho_A(0)>0\iff \underbrace{\thresh-\delta\cdot\cmax}_{ \text{Biomass produced in phase 1}}>\underbrace{(e^{r_{B,1}\tau_1} - 1 )}_{ \text{Net Growth of B in $\tau_1$}}\cdot\underbrace{{\delta \cmax}}_{ \text{Max population of B at start of phase 1}} \label{low_bound}
\end{equation}
And similarly, requiring that $\rho_B>0$, we have that
\begin{equation}
 \rho_B(0)>0\iff \underbrace{\thresh-\delta\cdot\cmax}_{ \text{Biomass produced in phase 1}}<\underbrace{(e^{r_{A,1}\tau_1} - 1 )}_ {\text{Net Growth of A in $\tau_1$}}\cdot\underbrace{{\delta \cmax}}_{ \text{Max population of A at start of phase 1}} \label{high_bound}
\end{equation}
Or by looking at phase 2,
\begin{equation}
 \rho_B(0)>0\iff \underbrace{\cmax-\thresh}_{\text{Biomass produced in phase 2}}>\underbrace{(e^{r_{A,2}\tau_2} - 1 )}_{ \text{Net Growth of A in $\tau_2$}}\cdot\underbrace{{\delta \cmax}\cdot e^{r_{A,1}\tau_1}}_{ \text{Max population of A at start of phase 2}} \label{high_bound2}
\end{equation}
But \eqref{low_bound} and \eqref{high_bound} are just saying that resource sharing must be possible for the necessary $\tau_1$ as we require that the resources consumed by the co-culture in $\tau_1$ be greater than the resources that a monoculture of $B$ would have consumed, and less what a monoculture of $A$ would have consumed (we need the second requirement as otherwise you do reach phase 2 before $\tau_1$, and thus you cannot have coexistence). Similarly, we get that the resources in phase 2 should be more than the monoculture of A could consume on its own. In fact, if we add up the two conditions, we get that you need resource sharing of total biomass must be possible). We note that both \eqref{low_bound} and \eqref{high_bound} are not necessarily true (trivially, we can move $\cmax-\thresh$ so that either \eqref{low_bound} or \eqref{high_bound} are not satisfied as the RHS is independent of $\cmax-\thresh$). Thus, only certain monocultures permit resource sharing. Further, we note that the ones that do, are stable since if there's resources that could be consumed that a monoculture cannot consume, it allows for invasion that can grow in a time period given by matrix inversion.\par
We now solve for $\rho_A(0)$ and $\rho_B(0)$ that satisfy Eq.~\textbf{\ref{r_tau}-\ref{res_2}}. First, from \eqref{r_tau}, we obtain, 
\begin{align}
    \tau_1&=-\log\delta\cdot\left(\frac{r_{A,2} -r_{B,2}}{r_{B,1}\cdot r_{A,2} -r_{B,2}\cdot {r_{A,1}}}\right)\\
    &=\log(1/\delta)\cdot \frac1{r_{B,1}}\left(\frac{p_B-1}{p_A\cdot p_B-1}\right).
\end{align}
And similarly, we obtain
\begin{align}
    \tau_2=-\log\delta\cdot\left(\frac{r_{A,1} -r_{B,1}}{r_{A,1}\cdot r_{B,2} -r_{A,2}\cdot r_{B,1}}\right)=\log(1/\delta)\cdot\frac{1}{r_{A,2}}\left(\frac{p_A-1}{p_A\cdot p_B-1}\right).
\end{align}
We note that this value is independent of $\thresh$ and $\cmax$. To obtain the steady-state values of $\rho_A(0)$ and $\rho_B(0)$, which we denote with $\rho_A^*$ and $\rho_B^*$ (note that $\rho_A^*+\rho_B^*=\delta\cdot\cmax$), we use Eq.~\textbf{\ref{res_1}-\ref{res_2}} to obatin
\begin{align}
    \thresh-\delta\cdot\cmax &= (e^{r_{A,1} \tau_1}-1) \rho_A^* + (e^{r_{B,1} \tau_1} - 1 )(\delta\cmax-\rho_A^*)\\
    \implies \rho_A^*&=\frac{\thresh-e^{r_{B,1} \tau_1}\delta\cmax}{e^{r_{A,1} \tau_1}-e^{r_{B,1} \tau_1}}\\
    &=\frac{\tthresh-\delta^{\frac{p_A-1}{p_A-1/p_B}}}{\delta^{\frac{1-p_B}{p_B-1/p_A}}-\delta^{\frac{1-p_B}{p_A\cdot p_B-1}}}\cdot \cmax,
\intertext{and similarly,}
    \rho_B^*&=\frac{\thresh-e^{r_{A,1} \tau_1}\delta\cmax}{e^{r_{B,1} \tau_1}-e^{r_{A,1} \tau_1}}\\&=\frac{\delta^{\frac{p_A-1}{p_A\cdot p_B-1}}-\tthresh}{\delta^{\frac{1-p_B}{p_B-1/p_A}}-\delta^{\frac{1-p_B}{p_A\cdot p_B-1}}}\cdot \cmax.
\end{align}
Thus, while there is a non-trivial dependence on $\delta$, $p_A$, and $p_B$, there is a linear dependence on $\tthresh$ and $\cmax$. Below, we explore the non-trivial dependence on $p_A$, and $p_B$. This analysis also tells us that there can only be one viable non-trivial solution for $\rho_A^*$ and $\rho_B^*$ (in other words, there can only be one non-trivial fixed point for the discrete map given by one growth-dilution cycle). This is why the complete system at steady-state can be understood by studying the trivial fixed points (i.e., the respective monocultures) as if both trivial fixed points are unstable to invasion, the system must have a non-trivial fixed point. Further, we can show that if it exists, the non-trivial fixed point must be stable. Intuitively, this is because if the frequency of Species A is increased, the amount of time it would take for the population to produce all of the allocated biomass in its preferred phase will be shorter, thus harming Species A. Similarly, the amount of time it would take for the population to accumulate all of the biomass in the phase that it is growing slower will be longer, once again harming Species A. Thus, there can only be one stable fixed point in the system.\par
We note that this method can be extended to many species and many phases as well. While it can be difficult to infer the final solution as it requires solving a system of transcendental equations of many variables, this approach does tell us that there can be exactly one solution for $\{\rho_\alpha^*\}$. This is because there is an unknown variable for each species ($\rho_\alpha^*$), and a corresponding equation (similar to \eqref{r_tau}) for each species. While this system of equations for each species features an unknown variable for each phase ($\tau_i$), the equations for resource constraints provide another set of equations for each $\tau_i$ (which is the transcendental system of equations). Thus, unless the system of transcendental equations produces a degeneracy for $\tau_i$, there is exactly one $\{\rho_\alpha^*\}$ that solves the system of equations.\par

\subsection{Negative frequency-dependent selection in growth-dilution cycles}
\label{lyapunov}
Here, we discuss the stabilizing mechanism for coexistence in growth-dilution cycles. 
We may understand this by considering only the dynamics of frequency of one species, say Species A, denoted by $x(t)$:
\begin{align}
    x(t)\equiv\frac{\rho_A(t)}{\rho_A(t)+\rho_B(t)}.
\end{align}
By changing variables in \eqref{base1}, we arrive at the following equation for the dynamics of $x$:
\begin{align}
    \dot{x}=(r_A-r_B)\cdot x\cdot (1-x).
\end{align}
This is the logistic growth equation with time-varying growth rates. This gives us that the frequency of Species A at the end of the cycle, $x(T)$, given that its frequency at the beginning of the cycle is $x_0$, is
\begin{align}
    x(T)=\frac{x_0}{x_0+(1-x_0)e^{-q(x_0)}}\label{deltax},
\end{align}
where $q(x_0)$ is the average fitness of Species A over Species B over one cycle:
\begin{align}
    q(x_0)=\int_0^T(r_A(s(t))-r_B(s(t)))dt.
\end{align}
This shows that $x(T)$ is uniquely determined by $x_0$ as $s(t)$ can be uniquely determined by solving the initial value problem given in \eqref{base1}-\eqref{base3}. Further, from \eqref{deltax} we note that the change in frequency of A, $x(T)-x_0$, has the same sign as $q(x_0)$, and $x(T)=x_0\equiv x^*$ if $q(x^*)=0$. This just reiterates that the frequency of A will be determined by its average relative fitness over one cycle. Such a point, $x^*$ would be a fixed point of the map that describes how the frequency changes over one cycle, $x_0\to x(T)$, since for $x=x^*$, the frequency doesn't change.\par
Thus, the dynamical behavior of the growth-dilution cycle is determined by $q(x_0)$. While determining $q$ is very difficult as we cannot know $s(t)$ without solving the dynamical system given by \eqref{base1}-\eqref{base3}, we can show that $q'<0$ for all $x_0$ (Section \ref{proof}). This shows that $q=0$ for at most one value of $x_0$ and allows us to construct a Lyapunov function,
\begin{align}
    V(x)=\int_x^{x^*} q(x')dx'.
\end{align}
If no $x^*$ exists such that $q(x^*)=0$, $x^*$ can be taken to the value closest to 0 (which is necessarily unique as $q'<0$). As verified in Section \ref{proof}, $V(x)$ fits all the requisite criteria for a discrete-time strict Lyapunov function that leads to global asymptotic stability. Thus, for any two growth dependences, the system always relaxes to the same steady state.\par

This result that $q'<0$ can be restated as saying that the average fitness of a species has a negative frequency dependence in growth-dilution cycles. This result holds for all $r_A$ and $r_B$. Further, the magnitude of $q'$ reveals how stable the system is. If $q'$ is very large, then the system quickly converges to the fixed point (Fig.~3G). A greater negative value of $q'$ indicates that the difference in growth rates is much larger than the mean growth rate. We note that this is the case when the growth dependences are highly non-linear.\par 

\subsubsection{Stabilization of Resource Trajectories}
A different perspective on the stabilizing nature of resource sharing can be obtained by looking at the trajectory that the co-culture traverses in the space of environmental variables. The key feature in the consumer-resource models that we use is that the trajectory is determined by the growth of the species. This is because the populations affect the environment through growth (either by consuming resources or secreting pollutants). Let us consider perturbations to the resource trajectory corresponding to a fixed point. Higher growth during a section of the trajectory means faster movement along the trajectory and less time spent at that section. But growth is also proportional to the time spent along the resource trajectory. Conversely, slower growth means more time at that section and thus the initial perturbation is countered. This leads to negative feedback for growth and, as a result, for resource change, thus stabilizing the resource trajectory. For the case of many environmental variables, movement along the trajectory can be projected onto the axis corresponding to each environmental variable, and the trajectory along each environmental variable can be considered independently. As movement along each projection is stabilized independently (such that at every point in time, there is a unique stable value of each environmental coordinate), the entire trajectory is stabilized.\par A similar understanding can be obtained by considering perturbations in the resource trajectory itself rather than perturbations in growth as we did above. A perturbation in the resource trajectory will either slow down or speed up growth, and this perturbation in growth will counteract the original perturbation in the resource trajectory.\par
However, this stabilization is local along every point on the resource trajectory, while coexistence is a global property of the entire trajectory. In other words, a negative frequency dependence of fitness does not mean that fitness ever has to be negative. Coexistence requires that each species have a negative fitness over the other species for some frequency, especially when it is abundant. This leads us to a simple necessary and sufficient criterion for coexistence: mutual invasibility (discussed in Section \ref{mic_derive}.
\subsubsection{Proof of Negative Frequency Dependence}
\label{proof}
Consider the frequency of Species A,
\begin{align}
    x_A=\frac{\rho_A}{\sum_\alpha \rho_\alpha}
\end{align}
Taking the time derivative, we have that
\begin{align}
    \dot{x_A}&=\frac{\dot{\rho_A}\cdot \sum_\alpha \rho_\alpha-{\rho_A}\cdot\sum_\alpha \dot{\rho_\alpha}}{(\sum_\alpha \rho_\alpha)^2}\\
    &=\frac{r_A\cdot {\rho_A}\cdot \sum_\alpha {\rho_\alpha}-{\rho_A}\cdot\sum_\alpha r_A{\rho_\alpha}}{(\sum_\alpha \rho_\alpha)^2}\\
    &=\frac{r_A\cdot {\rho_A}\cdot \sum_{\alpha\neq A} {\rho_\alpha}-{\rho_A}\cdot\sum_{\alpha\neq A} r_\alpha{\rho_\alpha}}{(\sum_\alpha \rho_\alpha)^2}\\
    &=\frac{r_A\cdot {\rho_A}\cdot \sum_{\alpha\neq A} {\rho_\alpha}-\bar{r}\cdot {\rho_A}\cdot\sum_{\alpha\neq A} {\rho_\alpha}}{(\sum_\alpha \rho_\alpha)^2}\\
    \text{where }&\bar{r}\equiv \frac{\sum_{\alpha\neq A} r_\alpha{\rho_\alpha}}{\sum_{\alpha\neq A} {\rho_\alpha}}=\frac{\sum_{\alpha\neq A}r_\alpha\cdot x_\alpha}{1-x_A}.\\
    &=(r_A-\bar{r})\cdot x_A \cdot (1-x_A)
    \end{align}
    Thus, for any time $t$, we have that
    \begin{align}
        x_A(t)=\frac{x_A^0}{x_A^0+(1-x_A^0)e^{-q(x_A^0,t)}}
    \end{align}
    where $x_A^0=x(t=0)$ and
    \begin{align}
        q_A(\{x_A^0\},t)=\int_0^t (r_A-\bar{r}) dt
    \end{align}
    This gives us that
    \begin{align}
        x_A(T)=\frac{x_A^0}{x_A^0+(1-x_A^0)e^{-q_A(x_A^0,T)}}.
    \end{align}
    Since this calculation holds for all species, we have a map from the species population at the beginning of the cycle to the species population at the end of the cycle. We define the Lyapunov function, $V$, such that
    \begin{align}
        V_A(\{x_\alpha(0)\})\equiv \int_{\{x_\alpha(0)\}}^{\{x_\alpha^*\}} q_A(\{x_A\},T)\cdot d(\{x_A\}).
    \end{align}
    For discrete time dynamics, the requirement for global convergence to $\{x_\alpha^*\}$ is that $V_A(\{x_\alpha^*\})=0$ (by definition), $V_A(\{x_\alpha(0)\})>0$, and $V_A(\{x_\alpha(T)\})-V_A(\{x_\alpha(0)\})<0$~\cite{bof2018lyapunov}.
    We note that
    \begin{align}
        V_A(\{x_\alpha(T)\})-V_A(\{x_\alpha(0)\})=-\int_{\{x_\alpha(0)\}}^{\{x_\alpha(T)\}} q_A(\{x_A\},T)\cdot d(\{x_A\})
    \end{align}
    But $d(\{x_A\})$ has the opposite sign as $q_A(\{x_A\},T)$ (as if $x_A(T)>x_A(0)$, then $q_A>0$) and thus the last requirement is also satisfied. All that is left for us to demonstrate convergence is to show that $V$ is positive. We do so by showing that $\nabla^2 V<0$. This requires that
    \begin{align}
        \frac{\partial q_A(\{x_A\},T)}{\partial x_A^0}&<0\\
        \iff \frac{\partial q_A(\{x_A\},T)}{\partial x_A}\frac{\partial x_A}{\partial x_A^0}&<0
   \end{align}
   But $\frac{\partial x_A}{\partial x_A^0}>0$ if $\frac{\partial q_A(\{x_A\},T)}{\partial x_A}<0$ so it suffices to show that 
   \begin{align}
   \frac{\partial q_A(\{x_A\},T)}{\partial x_A}&<0\\
        \iff 0>&\int_{s(0)}^{s(t)} \frac{\partial }{\partial x_A} \frac{r_A(s)-\bar{r}(x,s)}{ds/dt} ds\\
        \iff 0>&\int^{s(0)}_{s(t)} \frac{\partial }{\partial x_A} \frac{r_A-\bar{r}}{\sum r_\alpha\rho_\alpha} ds\\
        \iff 0>&\int^{s(0)}_{s(t)} \frac{-\bar{r}'\cdot (\sum r_\alpha\rho_\alpha)-(r_A-\bar{r})(\sum r_\alpha\rho_\alpha)'}{(\sum r_\alpha\rho_\alpha)^2} ds\\
        \iff 0>&\int^{s(0)}_{s(t)} \frac{-\bar{r}'\cdot (r_A\cdot x+\bar{r}\cdot (1-x))-(r_A-\bar{r})( r_A\cdot x+\bar{r}\cdot (1-x))'}{(\sum r_\alpha\rho_\alpha)^2/(\sum \rho_\alpha)} ds\\
        \iff 0>&\int^{s(0)}_{s(t)} \frac{\bar{r}'\cdot (r_A\cdot x+\bar{r}\cdot (1-x))+(r_A-\bar{r})( r_A+\bar{r}'\cdot (1-x)-\bar{r})}{(\sum r_\alpha\rho_\alpha)^2/(\sum \rho_\alpha)} ds
    \end{align}
    For the case of two species, $\bar{r}'=r_B'=0$. Thus, the condition is satisfied. 

\subsubsection{Unique Fixed Point for Discretized Growth Functions}
\label{discrete}
Assume that the range of $\rhotot$ can be discretized into $N$ niches such that each species $\alpha$ (out of $M$ total species) has a constant growth rate $r_{\alpha,n}$ in niche $n<N$. If there is a stable steady state, then
\begin{equation}
    \sum_n^N r_{\alpha,n}\tau_n=-\log \delta,\ \forall \alpha.\label{tau_n}
\end{equation}
where $\tau_n$ is the time spent by the community in niche $n$ in the steady cycle. Further, for each niche $n$, we have the following constraint:
\begin{equation}
    \eta_n=\sum_\alpha \rho_\alpha^*(0)\exp(\sum_{i<n}r_{\alpha,i}\tau_i)(\exp(r_{\alpha,n}\tau_n)-1),\label{eta_n}
\end{equation}
where $\rho_\alpha^*(0)$ is the steady cycle population density of species $\alpha$ at the beginning of the cycle.
\eqref{tau_n} and \eqref{eta_n} thus give us $N+M$ constraints for the $N+M$ unknown variables ($tau_n$ and $\rho_\alpha^*(0)$). Thus, unless there is a strong degeneracy such that the transcendental equation in \eqref{eta_n} yields multiple possible solutions (for example, if $r_{\alpha,n}$ is the same for all $\alpha$ in each niche $n$), there can only be one solution to the system of equations \eqref{tau_n} and \eqref{eta_n}.

\subsubsection{Existence of non-trivial solutions can be inferred by mutual invasibility}
Inferring when a possible frequency exists such that $q=0$ can be very complicated because the time dependence of the growth rates cannot be inferred without solving the ODEs numerically. However, because we know there is a negative frequency dependence, we can infer if such an $x^*$ exists by asking what the values of $q(0)$ and $q(1)$ are. If $q>0$ for all $x$, that would mean that Species A always has a fitness advantage over Species B. Similarly, if $q<0$ for all $x$, that would mean that Species B always has a fitness advantage over Species A.\par

The ability to understand the necessary and sufficient conditions for coexistence by considering only the cases that either species is present (as $x=0$ is the case when only Species B is present, and $x=0$ is the case when only Species A is present) is very valuable when studying systems with non-linear growth rates. The study of these cases, also known as invasion analysis, is ordinarily sufficient to demonstrate coexistence. But because of the result that $q'<0$, it is also necessary.\par
For our case, mutual invasibility can be verified without any numerical simulations. This is because the condition $q=0$ can be written as $\mathbb{E}(r_A/r_B)>1$, where $\mathbb{E}(\cdot)$ is the time-averaged expectation value. Mathematically, $\mathbb{E}(\cdot)$ is the integral of the argument from the lowest biomass value to the highest biomass value attained in the monoculture, weighted by $\omega(s)\equiv [-\log\delta \cdot(\frac{n_0}{1-\delta}-s)]^{-1}$. $\omega(s)$ is the equivalent of the partition function for biomass accumulation.\par
Similarly, $q(1)<0$ can we written as $\mathbb{E}(r_B/r_A)>1$. If both monocultures are invadable, then there necessarily is coexistence.\footnote{This also allows us to see that there cannot be bistability, as that would require that $\mathbb{E}(r_A/r_B)<1$ and $\mathbb{E}(r_B/r_A)<1$ but $\mathbb{E}(r_A/r_B)+\mathbb{E}(r_B/r_A)>2$ for all $r_B$ and $r_A$ as $x+1/x>2$ for all $x>0$.}\par
This criterion provides a general intuition of when there is coexistence: when both $r_A/r_B$ and $r_B/r_A$ are large for different parts of the growth step, both monocultures are invadable and there is coexistence. Thus, we need a range of biomass values when A grows much faster than B, and a range when B grows much faster than A. Various trade-offs can facilitate such a separation of biomass ranges, as does the strength of the non-linearity. This indicates to us why coexistence is increased when biologically-motivated modifications are introduced.\par
This criterion also indicates when interactions between two species can facilitate coexistence: when the interactions between two species lead to the initially slower-growing species eventually growing faster than the initially faster-growing species. This is highlighted in Amarnath et al.~\cite{amarnath2021stress} showing how stress-induced cross-feeding leads to stable coexistence in a marine co-culture. Species A grows faster than Species B initially, but then pollutes its environment by secreting acetate as a by-product of growth. This pollution leads to the supression of its own growth and the leakage of metabolites. Species B can thus consume these metabolites to grow. Thus, this leads to the creation of periods when A grows much faster than B, and a period when B grows much faster than A. Though the experiment was performed in growth-dilution cycles, in natural settings the periodic supply of food describes a very similar dynamic in the system. The experimental result was surprising because if the coexistence were due to the standard picture of commensal/mutualistic cross-feeding where one species steadily secretes a metabolite consumed by the other, there would be no coexistence for this pair of species as under ideal conditions, one species always grows faster than the other.\par
Thus, this mutual invasibility criterion provides a counter-point to the standard conception of coexistence due to mutualism in which both species promote each others' growth. In time varying environments, both species can end up limiting their own growth rates and thus coexist. This suggests that self-organized mechanisms by which species inhibit their own growth may facilitate coexistence.
\subsection{Mutual invasibility criterion}
\label{mic_derive}
Here, we derive the criterion presented in Eq.~\textbf{3}.
Let's consider a monoculture of Species A with a minimal amount of Species B such that $(\rho_A^{(j)}(0),\rho_B(0))=(\rho_A^0,\epsilon)$ and $\rho_A^0\gg\epsilon$. Thus,
\begin{align}
    \log\frac{\rho_B^{(j+1)}(0)}{\rho_B^{(j)}(0)}&=\int_{0}^{T}r_B(\rhotot(t))\cdot dt+\log\delta
    \intertext{Since $\rhotot$ is monotonic in $t$, we can substitute $t$ with $\rhotot$}
    &=\int_{0}^{T}\frac{r_B(\rhotot)\cdot d\rhotot}{d\rhotot/dt}+\log\delta\\
    &=\int_{\rho_A^0+\epsilon}^{\cmax}\frac{r_B(\rhotot)}{r_A(\rhotot)\cdot \rho_A^{(j)}(t)(\rhotot)+r_B(\rhotot)\cdot \rho_B^{(j)}(t)(\rhotot)}d\rhotot+\log\delta
    \intertext{Assuming $r_A(\rhotot)>0$, we can always choose $\epsilon$ such that $\rho_B^{(j)}(\rhotot)\ll  \rho_A^{(j)}(\rhotot),\ \forall\rhotot$, and thus $\rho_B^{(j)}(\rhotot)\ll  \rho_A^{(j)}(\rhotot)\cdot r_A(\rhotot)/r_B(\rhotot)$ and $\rhotot=\rho_A^{(j)}(\rhotot)$}
    &\approx\int_{\rho_A^0}^{\cmax}\frac{r_B(\rhotot)}{r_A(\rhotot)}\frac{d\rhotot}{\rhotot}+\log\delta.
    \end{align}
Similarly,
\begin{align}
    \log\frac{\rho_A^{(j+1)}(0)}{\rho_A^{(j)}(0)}&\approx\int_{\rho_A^0}^{\cmax}\frac{r_A(\rhotot)}{r_A(\rhotot)}\frac{d\rhotot}{\rhotot}+\log\delta=\log \cmax-\log \rho_A^0+\log \delta\\
    \implies \rho_A^{(j+1)}(0)&=\delta\cmax.
\end{align}
Thus, in subsequent cycles, 
\begin{align}
    \log\frac{\rho_B^{(j+k+1)}(0)}{\rho_B^{(j+k)}(0)}&\approx\int_{\delta\cmax}^{\cmax}\frac{r_B(\rhotot)}{r_A(\rhotot)}\frac{d\rhotot}{\rhotot}+\log\delta.\\
    \implies \log\frac{\rho_B^{(j+k+1)}(0)}{\rho_B^{(j)}(0)}&\approx k\int_{\delta\cmax}^{\cmax}\frac{r_B(\rhotot)}{r_A(\rhotot)}\frac{d\rhotot}{\rhotot}+\int_{\rho_A^0}^{\cmax}\frac{r_B(\rhotot)}{r_A(\rhotot)}\frac{d\rhotot}{\rhotot}+(k+1)\log\delta
\end{align}
For $k\to \infty$, 
\begin{equation}
\rho_B^{(j+k+1)}(0)\gg \epsilon\text{ if }I_{A,B}\equiv\int_{\delta\cmax}^{\cmax}\frac{r_B(\rhotot)}{r_A(\rhotot)}\frac{d\rhotot}{\rhotot}>\log \delta.
\end{equation}
This is the invasibility criterion in Eq.~\textbf{3}. If both $I_{A,B}>\log \delta$ and $I_{B,A}>\log \delta$, then there must be a non-trivial fixed point of the system, and by the proof in Section \ref{lyapunov}, it must be stable and further, the only stable solution.
\newpage
\subsection{Continuous Growth Relations}
\subsubsection{Monod Relation}
A popular choice in microbiology and ecology for $r_i$ is the Monod relation~\cite{monod1949growth}, also known as the Michaelis-Menten function or the Holling's type II functional response. It is chosen to describe growth that is proportional to the nutrient availability at low nutrient concentrations but saturates at high nutrient concentrations to a maximal rate.\par
We first consider the effect of varying the two environmental parameters, $s_0$ and $\delta$ for Species A and B with growth rates as described in Fig.~\ref{monod}A. We simulated six-hours long growth dilution cycles. In general, we find that the system is in the vicinity of a steady-state in less than $\sim100$ cycles (our results did not change significantly for longer cycles or for more cycles). The steady-state cycle is defined as having population densities and nutrient concentrations that are exactly the same in consecutive cycles.\par 
As can be seen in Fig.~\ref{monod}A for our choice of physiological parameters ($r_i^{\max}$ and $K_i$), Species A (shown in red) has a higher growth rate than Species B (shown in blue) when the nutrient concentration is high (because $r_A^{\max}>r_B^{\max}$), while Species B has a higher growth rate when the nutrient concentration is low (because $K_A^{\max}>K_B^{\max}$). Such a trade-off is known as the opportunist-gleaner trade-off~\cite{grover1997resource}. Although the empirical existence of such a trade-off is debated~\cite{kiorboe2020heterotrophic,letten2021gleaning,fink2022microbial}, we note that in this simplest case which has no other interactions, such a trade-off is necessary for coexistence as otherwise one species will always have a lower growth rate than the other and thus eventually be out-competed.\par 
In Fig.~\ref{monod}B, we report the average fitness (average difference in growth rate over one cycle) of Species A over Species B after 100 cycles for different environmental parameters. A positive fitness value (denoted by red shading) indicates that the population of Species A is driving the population of Species B down, and thus Species B will eventually be removed from the system. A negative fitness value (denoted by blue shading) indicates that the population of Species A is being driven down and will eventually be removed from the system, while a near-zero fitness value (denoted by white shading) means that both species have reached a non-zero steady state population and thus will coexist indefinitely. As can be seen and as would be expected, high nutrient supply favors the species with the higher value of $r_i^{\max}$, while low nutrient supply favors the species with the lower $K_i$. Similarly, a lower dilution factor favors the species with the higher value of $r_i^{\max}$ as the relative amount of time spent in higher nutrient concentrations is higher.\par
We also report the average fitness of Species A over Species B for different physiological parameters and fixed environmental parameters ($\delta=0.1, s_0=10 K_B$) in Fig.~\ref{monod}C. As would be expected, if there is no opportunist-gleaner trade-off, as in the top left and bottom right quadrants of the phase plot, there would be no coexistence. Further, even if there is a trade-off, coexistence is not guaranteed as can be seen in the other two quadrants which have red, white, and blue regions. Thus, coexistence is not a trivial consequence of the trade-off. However, there is a narrow parameter regime of coexistence between the region where Species A dominates and where Species B dominates.\par
In 1972, Stewart and Levin showed mathematically that two species with an opportunist-gleaner trade-off may coexist indefinitely in growth-dilution cycles (Fig.~1C).
They also demonstrated that such a coexistence was ``structurally stable'', i.e., it was robust to noise in the environmental/experimental parameters (Fig.~\ref{monod}B). As can be seen, this coexistence is also robust to small physiological perturbations (Fig.~\ref{monod}C) and there exists an entire region in the environmental and physiological parameter phase space (shaded in white) where the two species coexist, rather than just being a boundary between the two regions. This is a key result as it shows that this kind of coexistence is not a result of narrow fine-tuning, but seems to incorporate stabilizing mechanisms that promote coexistence. While this was a crucial novel result, it has been mostly dismissed in literature~\cite{grover1997resource}, due to the relatively small region of coexistence in parameter phase space. However, as we will see below, this region of coexistence is significantly broadened when biologically-realistic effects are considered. As such, it indicates a more general emergent principle for self-stabilized coexistence rather than serving as just a niche special case.\par

\subsubsection{Cut-off due to Maintenance Energy}
Bacterial species require a certain minimum amount of energy, known as maintenance energy, to be able to grow. If sufficient non-zero amount of nutrients to generate this energy are not supplied, the bacteria cease to grow. This effect can be incorporated by subtracting a constant value of maintenance energy consumption rate, given by $J_i$, from the growth rate of each species and setting the minimal growth rate to be 0 (i.e, we exclude death). This leads to the following form for the growth rate:
\begin{align}
    r_i(s)=r_i^{\max}\cdot\frac{s}{s+K_i}-J_i.
\end{align}
This expands the coexistence region of phase space considerably. In fact, for almost any two species defined by the physiological parameters $r_i$, $K_i$, and $J_i$, the coexistence region of the environmental phase-space occupies an entire half-plane (see Fig.~\ref{monod}E). This is because at very low resource concentrations, one of the species necessarily grows, while the other doesn't. The species that grows at very low resource concentrations will subsequently always survive for all environmental parameters.\par
The coexistence region in physiological parameter space is also extended, with species benefiting from even relatively minor differences in resource affinities. This is because strict coexistence only requires sufficient fold change between cycles rather than any minimal abundance. 

\subsubsection{Hysteretic Growth Kinetics}
We note that the Monod relation is obtained empirically for a perfectly-adapted bacterial population, i.e., the bacterial population is grown at a static nutrient concentration and its growth rate is measured for that nutrient concentration. In a growth-dilution cycle, the bacterial population may not have adapted to the nutrient concentration it experiences at any instant as the nutrient concentration is constantly changing. This can be viewed as a hysteresis in the growth-rate due to the slow adaptation of the internal state of the cells that constitute the population. Similar hysteretic effects are modeled by the Droop model which is popular in studying algae~\cite{wang2022mathematical} and is also seen in macroscopic organisms as a predator's searching, attacking, or handling efficiency often empirically increases as prey density increases. This is because the feeding response of organisms often display some form of learning behavior, as a predator must have a minimal encounters with its prey before the predator is maximally efficient at feeding on that prey item.\par Such a growth dependence is known as a Type III functional response and is described by the Hill function~\eqref{hill},
\begin{align}
    r_i(s)=r_i^{\max}\cdot\frac{s^k}{s^k+(K_i)^k},\label{hill}
\end{align}
where $k>1$ is a positive number that denotes the number of minimal encounters the consumer must have with its food to increase its efficiency~\cite{real1977kinetics}.\par 
We note that the Monod relation is a special case of the Hill function (when $k=1$), and increasing $k$ leads to an increase in the coexistence region of phase-space (see the case of $k=2$ shown in Fig.~\ref{monod}D-F).\par%
\subsubsection{Lag Time}
Another effect that can be incorporated is the presence of a lag-time, such that both species do not grow for a short period of time at the beginning of each cycle. By design, the presence of lag times benefits the species with the shorter lag-time. This does not significantly increase (or decrease) the  coexistence region of the physiological and environmental phase spaces but rather shifts it in favor of one species. We note that though previous studies~\cite{manhart2018trade} found coexistence and bistability for species with constant growth rates and lag times in growth-dilution cycles, these studies require the dilution step to be a bottleneck step such that the total initial population is fixed (which requires changing the dilution amount every cycle). These studies also required very large differences in the yield (of the factor of $\sim$100-1000) between the two species. In the absence of these two effects, the coexistence disappears for constant growth rates~\cite{lin2020evolution} but persists for nonlinear growth rates.\par
A result that has been indicated numerically by previous studies for the case of Monod growth~\cite{stewart1973partitioning,wortel2023evolutionary,smith2011bacterial} and is implicit in the phase diagrams of Fig.~\ref{monod} is that there is no initial condition dependence in this system. Thus, there are only three steady-state possibilities that are determined by environmental and physiological parameters alone and hold for all initial conditions: that Species A eventually takes over the system, that Species B eventually does so, or that both species coexist indefinitely in a defined steady-state cycle. In other words, there is no bistability or multistability. We also note from Fig.~\ref{monod} that coexistence is determined by all six parameters that effectively describe the system and that each parameter can be varied to lead to any of the three possible outcomes. This implies that the system cannot be described as being a trivial outcome of a simple characteristic of the environment or physiology of the two species, but requires an interplay of the environment and the physiologies. Thus, though it is necessary that the growth rate dependences of the two species intersect, it is far from sufficient.\par

We note that if all $r_i$ are linear in the concentration of the nutrient (also known as a Holling's type I functional response), coexistence is not possible and only one species can survive \cite{armstrong1980competitive}. This is because one species will always grow faster than the other species and thus out-compete the other species over many cycles. Thus, nonlinear growth dependences on nutrient concentrations are required for coexistence and reflect a trade-off in the growth process.\par

Thus, we note that even for two species competing for a single nutrient, the coexistence region of parameter phase space is not necessarily small, but possibly a result of multiple physiological trade-offs between growth rate, resource affinity, adaptation time scale, and survivability in low resource conditions. For other systems shown in Fig.~1, the trade-offs could be due to susceptibility to a pollutant, cross-feeding, or anomalous response to environmental stress. Accordingly, growth-dilution cycles result in a consumer-resource analog to winnerless competition models, where transient dynamics enable multiple species to persist~\cite{afraimovich2008winnerless}.

\subsection{Derivation of Equi-abundance Solution}
\label{equi_derive}
We take each species, $\alpha$, to have growth rate $r_-$ basally and $r_+$ in its preferred growth rate. In the steady cycle, we look for the solution where there are $N$ species, each with abundance $1/N$. First, since this is the steady cycle, we require
\begin{align}
    \sum_i r_{\alpha,i}\cdot \tau_i=\log(1/\delta),
\end{align}
where $\tau_i$ is the time spent passing through each niche $i$. Plugging in the growth rates, we have
\begin{align}
    r_+\tau_\alpha+r_-(T-\tau_+)=\log(1/\delta),
\end{align}
where $\tau_\alpha$ is the time spent in the preferred niche. Since the rest of the equation carries no parameters unique to species $\alpha$, $\tau_\alpha$ must be the same for all species. Thus, we have that
\begin{align}
    \tau\equiv\tau_\alpha=\frac{\log(1/\delta)}{r_++r_-(N-1)},\ \forall \alpha
\end{align}
Now, we seek to find $\Delta \eta_n$ for each niche. Since this is the biomass consumed by all species in that niche, we have
\begin{align}
    \Delta \eta_n=\underbrace{\sum_{i<n}\frac{1}{N}\exp((n-2)r_-\tau+r_+\tau)(\exp(r_-\tau)-1)}_{\text{Species that have a preferred niche before the nth niche}}&+\underbrace{\frac{1}{N}\exp((n-1)r_-\tau)(\exp(r_+\tau)-1)}_{\text{Species that prefers the nth niche}}\\&+\underbrace{\sum_{i>n}\frac{1}{N}\exp((n-1)r_-\tau)(\exp(r_-\tau)-1)}_{\text{Species that have a preferred niche after the nth niche}}
\end{align}
After some simplification,
\begin{align}
    \Delta \eta_n&=\frac{\exp((n-1)r_-\tau)(\exp(r_-\tau)-1)}{N}\left(\sum_{i<n}\exp((r_+-r_-)\tau)+\frac{\exp(r_+\tau)-1}{\exp(r_-\tau)-1}+\sum_{i>n}1\right)\\
    \implies \Delta \eta_n&=\frac{\exp((n-1)r_-\tau)(\exp(r_-\tau)-1)}{N}\left((n-1)\exp((r_+-r_-)\tau)+\frac{\exp(r_+\tau)-1}{\exp(r_-\tau)-1}+N-n\right)
    \label{eq:equi-abundance_si}
\end{align}
In the case that the correction term is independent of $n$, i.e.,
\begin{equation}
n(\exp((r_+-r_-)\tau)-1)\ll\frac{\exp(r_+\tau)-1}{\exp(r_-\tau)-1}+N-\exp((r_+-r_-)\tau), \label{correction}
\end{equation}
we have that 
\begin{align}
    \Delta \eta_n&\propto\exp(nr_-\tau)\implies \Delta \eta_n\sim \delta^{\frac{n}{p+N-1}}
\end{align}
and the niche widths must be exponentially spaced. We now explore the cases where this is true.
This is obviously the case if $r_+=r_-$ as then $\exp((r_+-r_-)\tau)-1=0$.\\
In the case that $N\gg p\equiv \frac{r_+}{r_-}>1$, we have that
\begin{equation}
    \exp(r\tau)=\exp(\frac{r/r_-\cdot \log(1/\delta)}{p+N-1})\approx 1+\frac{r/r_-\log(1/\delta)}{N}
\end{equation}
Thus, the LHS in correction term is given by
\begin{equation}
    n\left(\frac{p-1-\log(1/\delta)}{N}\right)\ll N
\end{equation}
while the RHS is given by
\begin{equation}
\frac{\frac{p-\log(1/\delta)}{N}}{\frac{1-\log(1/\delta)}{N}}+N-\frac{p-1-\log(1/\delta)}{N}-1 \approx N.
\end{equation}
Thus, \eqref{correction} is satisfied.
Also, in the case that $p\gg N>1$, 
\begin{equation}
    \exp((r_+-r_-)\tau\to \exp(r_+\tau)=\exp\left(\frac{\log(1/\delta)}{1+\frac{N-1}{p}}\right)\approx \frac{1}{\delta}.
\end{equation}
and 
\begin{equation}
    \exp(r_-\cdot \tau\to \exp\left(\frac{\log(1/\delta)}{p+N-1}\right)\approx 1+\frac{\log(1/\delta)}{p+N-1}.
\end{equation}
Thus, 
\begin{align}
\Delta\eta_n&=\frac{\exp((n-1)r_-\tau)}{N}\frac{\log(1/\delta)}{p+N-1}\left(n(1/\delta-1)+\frac{\frac{1}{\delta}-1}{\frac{\log(1/\delta)}{p+N-1}}+N-\frac{1}{\delta}\right)\\
&=\frac{\exp((n-1)r_-\tau)}{N}\log(1/\delta)\left(\frac{n(1/\delta-1)}{p+N-1}+\frac{\frac{1}{\delta}-1}{\log(1/\delta)}+\frac{N-\frac{1}{\delta}}{p+N-1}\right)\\
&\approx \frac{\exp((n-1)r_-\tau)}{N}\left(\frac{1}{\delta}-1\right)\propto\delta^{\frac{n}{p+N-1}}
\end{align}

\subsection{The diagonal preference model} 
Based on the equiabundance solution, we consider a ``diagonal model'' in growth preference, in which each niche $n$ has a dominant species $\hat{\alpha}(n)$ whose growth rate $r_{\hat{\alpha}(n),n}$ well exceeds the growth rate of other species in that niche, and each species is dominant only in one unique niche. 

We anticipate the equi-abundance case to be highly stable. We validated this expectation by creating ensembles of $r_{\alpha,n}$ and $\Delta \eta_n$ values fluctuating within $\pm 30\%$. Notably, every species persists in $\sim 80\%$ of scenarios with near equi-abundance (Fig.~S6B, S6C). We also found that higher growth preference $p$ species in its main niche (with the accompanying change in the exponential niche distribution) increased survival rates of over 95\%. In fact, even with just a 3-fold growth preference and $\pm 30\%$ noise, all species remained in 45-65\% of cases. This nonlinear reliance on growth preference indicates diminishing returns. 

Yet, as Fig.~S6D illustrates, the proportion of preserved species for a set growth preference ($p$) declines as species and niche counts rise. The drop depends on $N/p$ (Fig.~S6E), emphasizing enhanced competition from the growing collective of slow-growing species.

Alternatively, the priority effect can be overcome by having the species preferred in earlier niches take on reduced growth preferences but equal niche widths. 
We find that exponentially-distributed growth rates for species in their preferred niches, 
\begin{equation}
r_+(n) \propto (1/\delta)^{n/(N+p-1)}, \label{r-eta}
\end{equation}
with $r_-$ fixed, retains a similar diversity (Fig.~S6G, S6H) as for exponentially distributed niche widths (Fig.~S6B, S6C).
See Fig.~S6. 

Collectively, these results underscore the tremendous (exponential) growth advantage of species specializing in early niches if $p$ is not too large, and hence the much higher relative dominance required for species specializing in the late niches to be maintained in the community.\par

\subsection{Resource Sharing in Consumer-Resource Models in a Chemostat}
One might interpret the states in the Community State (CS) Model as analogous to resources in a Consumer-Resource model. In the case of large communities, it leads to the question of how many species can be expected to coexist if $N$ species shared $N$ resources, with each species growing on multiple resources.

In the context of the CS Model, species coexist by occupying distinct niches, with niche width and overlap playing critical roles in determining community structure. Translating this to the CR model framework, we postulated that the allocation and consumption of resources could be a surrogate for these niche dynamics. Thus, we simulated the CR model in a chemostat. Each species was described by a consumption/growth matrix with diagonal elements set to 1 (such that each species invariably consumed a specific resource) and off-diagonal elements set randomly as 1 or 0 with the constraints described below. The growth rates,  We considered two distinct cases:

$K_s=M$ Case: In this case, we fixed the number of species that show rapid growth per resource. This setup parallels a situation in the CS Model where each niche supports a similar number of species. The results can be seen in Supplementary Figure S7.

$K_n=M$ Case: In this case, we explored a scenario where each species is allocated a fixed number of resources, akin to each species in the CS Model occupying $M$ niches. The results can be seen in Supplementary Figure S8.

\printbibliography